\documentclass[paper,nofootinbib,noshowkeys,superscriptaddress]{revtex4}
\usepackage[latin1]{inputenc}
\usepackage{pstricks,pst-node,pst-text,pst-3d,pslatex}
\usepackage[T1]{fontenc}
\usepackage{amsmath}
\usepackage{amsthm}
\usepackage{pgfplots}
\usetikzlibrary{matrix}
\usepackage{subfigure}
\usepackage{tikz}
\usepackage[font=small,labelfont=bf,
   justification=justified,
   format=plain]{caption}
\usetikzlibrary{automata,arrows,positioning,calc}
\usetikzlibrary{decorations.markings}

\pgfplotsset{/pgf/number format/use comma,compat=newest}

\usepackage{graphicx,amsfonts}
\usepackage{amsmath}
\usepackage{amsthm}
\usepackage{color}

\newcommand{\be}{\begin{eqnarray}}
\newcommand{\ee}{\end{eqnarray}}
\newcommand{\la}{\langle}
\newcommand{\ra}{\rangle}

\newcommand{\Id}{\mathbb{I}}
\newcommand{\nn}{\nonumber}

\theoremstyle{plain}
\newtheorem{teo}{Theorem}[section]
\newtheorem{prop}[teo]{Proposition}

\begin{document}
\author{S.~Orioli}
\affiliation{Dipartimento di Fisica, Universit\`a degli Studi di Trento, Via Sommarive 14 Povo (Trento), 38123 Italy. }
\affiliation{Trento Institute for Fundamental Physics and Applications (INFN-TIFPA), Via Sommarive 14 Povo (Trento), 38100 Italy. }
\author{P. Faccioli}
\affiliation{Dipartimento di Fisica, Universit\`a degli Studi di Trento, Via Sommarive 14 Povo (Trento), 38123 Italy. }
\affiliation{Trento Institute for Fundamental Physics and Applications (INFN-TIFPA), Via Sommarive 14 Povo (Trento), 38100 Italy. }

\title{ Dimensional Reduction of Markov State Models from  Renormalization Group Theory}

\begin{abstract}
 Renormalization Group (RG) theory provides the  theoretical framework to define Effective Theories (ETs), i.e. systematic low-resolution approximations of arbitrary   microscopic models. 
  Markov State Models (MSMs)  are shown to be rigorous  ETs for Molecular Dynamics (MD). Based on this fact, we use Real Space RG to vary the resolution of a MSM and define an algorithm for clustering  microstates into macrostates. The result is  a lower dimensional stochastic model which, by construction, provides the optimal coarse-grained Markovian representation of the system's  relaxation kinetics. To illustrate and validate our theory, we analyze a number of test systems of increasing complexity, ranging from synthetic  toy models to two realistic applications,  built form all-atom MD simulations.  The computational cost  of computing the low-dimensional model remains affordable on a desktop computer even for thousands of microstates.    
\end{abstract}
\maketitle
\section{Introduction}
MSMs -- for  recent reviews see \cite{Pande0, Noe0, Prinz0, Noe1}, for a discussion of mathematical aspects see \cite{MSMfund1,MSMfund2,MSMfund3, MSMfund4}--
  provide a coarse-grained description of  the conformational dynamics of macromolecules in which  the probability for the system to be found in a discrete set of metastable states (herby termed \emph{microstates}) evolves in time according to a master equation. 
The microstates and all the parameters of the master equation are obtained by reducing the data generated by MD simulations.

Coarse-graining the dynamics at the MSM level enables to extract the relevant kinetic and thermodynamical information from  many short independent MD trajectories.  The larger is the number of microstates in the model, the shorter are the MD trajectories which need to be run in order to specify the master equation. By choosing a sufficiently large number of microstates, the   computational effort can be massively distributed  and it becomes possible to investigate the dynamics for  time intervals  inaccessible to straightforward MD simulations.  For example, in applications to protein dynamics,  $10^2-10^3$~\cite{Pande1, Bowman0} microstates are usually required to achieve an efficient sampling of  the configuration space. 

Unfortunately, with such a large number of states it is hard to gain  insight into the conformational dynamics  by directly inspecting the transition pathways in the space of microstates. A way around this problem consists in further coarse-graining  the representation of the dynamics,  by clustering (lumping)  the  microstates into supergroups that are usually referred to as  \emph{macrostates}.  Physically, macrostates are to be interpreted as larger metastable regions of the configuration space,  which comprize many  smaller and  fast interconverting  microstates. 

Several algorithms  have been developed  to perform dimensional reduction of MSMs (see e.g. \cite{Huang0, Schutte0, Bowman0}), and some of these have been compared and assessed on a few benchmark systems in Ref.~\cite{Bowman1}. It was  found that the structure and the optimal number of macrostates is quite sensitive to the specific algorithm adopted to lump the microstates. Furthermore, all the considered algorithms  were found to fail  in reproducing the kinetics of the original MSM, by underextimating the relaxation timescales   by a factor as large as $\sim 20-100$. Alternative lumping schemes which provide relaxation times scales in much better agreement with the original model have been developed using Hidden Markov models \cite{hidden} or by means of projection techniques \cite{pcca, HummerSzabo, DeLosRios1, DeLosRios2}. 

In this work, we frame the theoretical foundation of MSMs and their dimensional reduction, within the ET formalism.
ETs --for an excellent pedagogical introduction see \cite{Lepage}--- are systematic low-resolution approximations of arbitrary more microscopic theories. Unlike  coarse-graining approaches inspired by heuristic or phenomenological arguments, ETs can be  rigorously derived starting from the underlying microscopic theory using RG theory. Consequently, under certain well-defined conditions to be discussed below, ETs are guaranteed to approximate the long-distance (or long-time)  dynamics within a degree of accuracy which can be estimated \emph{a priori} and systematically improved.

The physical picture behind  the ET formalism is very familiar: Any experimental probe with wavelength $\lambda$ (or frequency $\nu)$ is insensitive to the \emph{details} of the physics at lengthscales $\ll \lambda$ (or timescales  $\ll 1/\nu$). As a consequence, as long as one is interested on the  infra-red (IR) physics, i.e. in the behaviour of observables at distances $\gg \lambda$ or times $ \gg 1/\nu$,  all the so-called ultra-violet (UV) details of a microscopic theory are irrelevant, and can be accurately mimicked by a set of effective parameters. These parameters have to be computed from the underlying microscopic theory, or extracted from  experimental data.  Clearly, ETs are only 
applicable to physical systems which display a gap in characteristic scales, i.e. for which it is possible to separate IR from UV scales.

A familiar example of  ET is the multi-pole expansion of classical electrodynamics: The electric field ${\bf E}({\bf x})$ generated by an arbitrary localized charge distribution $\rho({\bf x})$  of  size $\sim \lambda$ is systematically approximated at distances $|{\bf x}| \gg \lambda$ by the series of multi-poles. In the terms of this series, the fine-grained structure of the charge distribution $\rho({\bf x})$ at the scale $\lambda$  is mimicked by effective constants like total charge, dipole moment, etc..
This example illustrates that ETs  are usually much simpler than the corresponding underlying microscopic theories. On the other hand, their  accuracy of any ET breaks down  at the scales comparable with the UV cut-off.


The identification of  MSMs with ETs paves the door to using RG methods in order to systematically lower their time- and space- resolution. 
In fact, one of the  main results of this  work is the derivation of a  physically sound and mathematically rigorous dimensional reduction scheme for MSMs based on  
the so-called Real-Space RG  formalism. This approach was inspired by the work of Degenhard and Rodriguez-Laguna \cite{RG1, RG2}, who used the RG  to lower the computational cost of integrating non-linear  partial differential equations.

In the following sections, we shall first review how MSMs emerge as rigorous ETs for MD. Then, in section \ref{RSRG} we develop our RG scheme for dimensional reduction and describe the practical implementation of the corresponding algorithm. In sections \ref{tests} and \ref{MDtest} we present a number of illustrative applications of increasing complexity  and assess the accuracy of  our  scheme. Finally, our main results are summarized in section \ref{conclusions}.

 \section{Theoretical Framework of Markov State Models}
\label{stochdiff}

MSMs can be defined without explicitly referring to a particular type of MD \cite{Sarich}. However, from a mathematical standpoint,  the connection between MSMs and rigorous ETs is particularly manifest  for systems which obey the over-damped Langevin equation. In turn, in view of  Zwanzig-Mori projection formalism \cite{Zwanzig0, Mori}, such a stochastic differential equation can be regarded as the  low-energy approximation of an underlying Hamiltonian dynamics.

Let us therefore consider a system composed of $N_a$ atoms, which obey the equation
\begin{equation}
\dot {\bf x}_i = -\frac{1}{k_B T} D~ \nabla_i U(x) + \eta_i(t),
\label{Lan2}
\end{equation}
where $x=({\bf x}_1,{\bf x}_2,...,{\bf x}_{N_a})$ is a point in  configuration space, $U(x)$ is the potential energy and $D$ is the diffusion coefficient  (assumed to be the same for all atoms, for the sake of notational simplicity). 
$\eta_j(t)$ is a white Gaussian noise obeying the fluctuation-dissipation relationship,
\be
\langle\eta_i(t)\cdot \eta_j(t')\rangle= 6 D~ \delta_{i j}~ \delta(t-t'), \qquad i,j=1, \ldots, N_a.
\ee

The probability distribution sampled by the stochastic differential Eq. (\ref{Lan2}) satisfies then the Fokker-Planck (FP) equation 
\be\label{FPE}
\frac{\partial}{\partial\,t}~P(x,t) = - H_{\text{FP}} P(x,t)
\ee
where $H_{\text{FP}}$  is the non-hermitian operator 
\be
\label{HFP}
H_{\text{FP}}= -D \sum_{i=1}^{N_a} \nabla_i\cdot \left(\nabla_i + \beta \nabla_i U(x)\right)\qquad (\beta \equiv 1/k_B T) \; ,
\ee
 
From $H_{\text{FP}}$ it is possible to define  a hermitian operator $H_{h}$ by performing the following non-unitary transformation:
\be\begin{cases}
\label{Heff}
H_{h} = e^{\frac{\beta}{2} U(x)}~ H_{\text{FP}} ~e^{-\frac{\beta}{2} U(x)} =- D \nabla^2 + \frac{D\beta^2}{4}~ \left[ \left( \nabla  U(x) \right)^2 - \frac{2}{\beta}\, \nabla^2 U(x)\right] \\
\label{psi}
\psi(x, t) = e^{-\frac{\beta}{2} U(x)} P(x,t)
\end{cases}
\ee
This transformation turns the FP equation into a Schr\"odinger equation in imaginary time:
\be
-\frac{\partial}{\partial t} \psi(x,t) = H_{h}~\psi(x,t).
\label{SE}
\ee
In the following, the function $\psi(x,t)$ will be referred to as the \emph{hermitian component} of the probability density $P(x,t)$. 

It is straightforward to show  that  $H_{h}$ and $H_{\text{FP}}$ have the same  spectrum, which is non-negative definite and contains a null eigenvalue,  $0= \lambda_1 < \lambda_2 < \ldots$. 
The  right zero-mode  $ f^R_1(x)$ is the Gibbs distribution:
\be
H_{\text{FP}} f^R_1(x) =  0,\qquad
 f^R_1(x) = \frac{1}{Z}~e^{-\beta U(x)},
\ee
where $Z=\int dx~e^{-\beta U(x)}$ is the system's canonical partition function.
The hermitian components of the left- and right- eigenstates of $H_{FP}$ are  related to the  eigenstates $\phi_i(x)$ of the corresponding hermitian  operator $H_{h}$:
\be
f^{R}_i(x) &=& e^{-\frac{\beta}{2} U(x)}~ \phi_i(x)\nn\\
f^{L}_i(x) &=& e^{\frac{\beta}{2} U(x)}~\phi_i(x) \;.
\ee
Finally, we note that the probability density $P(x,t)$ entering the FP equation (\ref{FPE}) can be expanded as a series of  right eigenfunctions of $H_{\text{FP}}$ or, equivalently,  eigenfunctions of $H_{h}$:
\be \label{time-dep-decomp}
P(x,t) = \sum_{i=1}^{\infty} c_i~f_i^R(x)~e^{- \lambda_i t}=   e^{-\frac{\beta}{2}U(x)}  \sum_{i=1}^{\infty} c_i~\phi_i(x)~e^{- \lambda_i t} \quad \text{where} \quad c_i \equiv \int dx f^L_i(x) P(x,0).
\ee

\subsection{Definition of Microstates}
\label{Markov states} 

\begin{figure}[t!]
\begin{center}
\includegraphics[width = 14 cm]{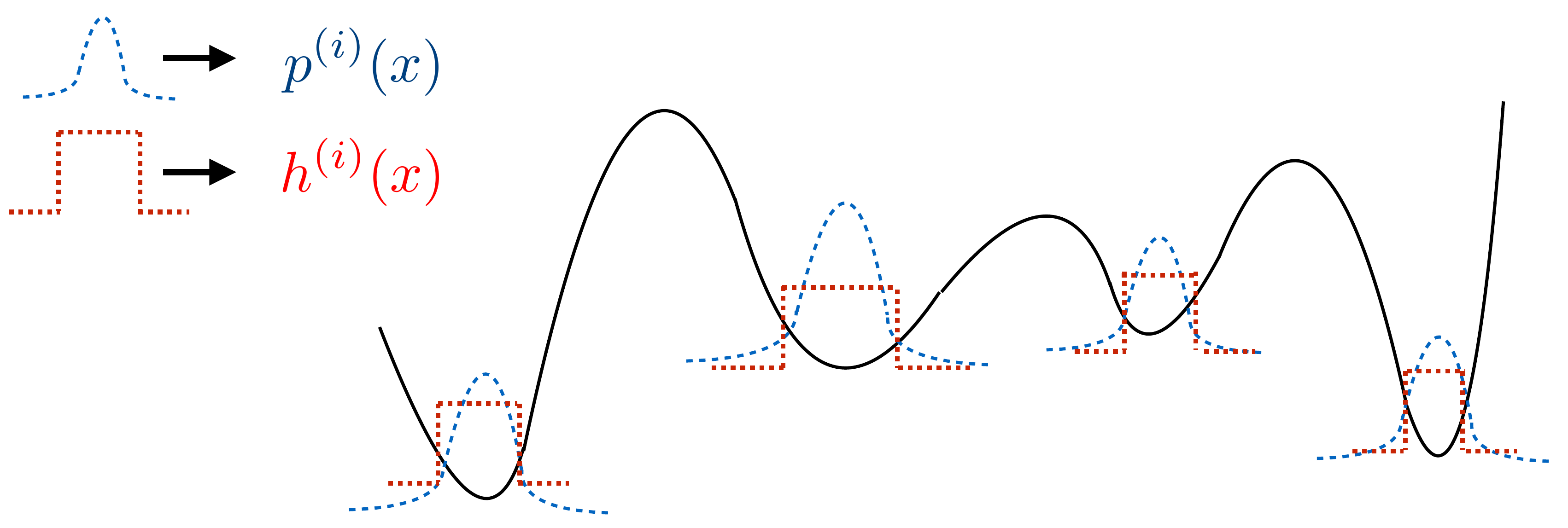}
\caption{Graphical representation of the definition of micro state distributions $P_i(x)$ and characteristic functions $h_i(x)$. The continuous line denotes the energy landscape. }
\label{Statefig}
\end{center}
\end{figure}

Let us now specialize on molecular systems in which  the typical relaxation times required to attain local thermal equilibrium within all  metastable states  are much smaller than the timescales associated to transitions between such metastable states. This condition is realized if the thermal energy $k_BT$ is much lower than the energy barriers between local minima and implies that the spectrum of  the $H_{\text{FP}}$ operator is gapped. In the following, we shall always assume such a low-temperature regime.
 Consequently, MSM can only  deal with dynamics at  time intervals much larger than the inverse of the  lowest eigenvalue above the gap, $\Lambda$. Thus, $\Lambda$ represents a typical value for the UV cut-off scale of the ET, i.e. $t\gg dt \equiv 1/\Lambda$. 

To explicitly construct such an ET, starting from the microscopic Fokker-Planck dynamics, we observe (see e.g. the discussion in Ref. \cite{MSMfund4}) that there exist exactly $N$ linear combinations  of the right-eigenstates  below the gap, 
 \be\label{Pdef}
p^{(i)}(x) = \sum_{j=1}^{N} C_{i j} ~f_{j}^R(x)
\ee
which simultaneously satisfy the following three properties: (i) non-negativity,  $p^{(i)}(x)\ge 0$, (ii) disjointness,  $p^{(i)}(x) p^{(j)}(x)= 0 $ for $i\ne j$ and $\forall x$ and (iii) local Gibbseanity,  
\be \label{local_gibbs}
p^{(i)}(x) =\frac{1}{z_i}e^{-\beta U(x)} h^{(i)}(x) \qquad \text{~with~ } \quad z_i = \int dx ~h^{(i)}(x)~ e^{-\beta U(x)},
\ee
where $h^{(i)}(x)$ is negligible everywhere and equal to 1 in the vicinity of one and only one of the local meta-stable states.
$p^{(i)}(x)$ is then interpreted as the probability distribution associated to the $i-$th so-called Markov state and  $h^{(i)}(x)$ denotes the corresponding characteristic function (see Fig. \ref{Statefig}).  

It is also relevant to consider the following linear combinations of the  left eigenstates $f^L_i(x)$, with $1 \le i\le N$:
\be\label{Adef}
a^{(i)}(x) =\sum_{j=1}^{N} C_{i j}~ f_L^j(x).
\ee 
The interest in these linear combinations resides in the fact that the $a^{(i)}(x)$ functions are proportional to the  characteristic functions of the microstates, thus are approximatively constant where $p^{(i)}(x)$ is non-negligible, and are negligible elsewhere. To see this, it is sufficient to isolate the hermitian component of the $a^{(i)}(x)$ distributions:
\be
a^{(i)}(x) = \sum_{j=1}^{N} C_{i j}~\phi_j(x)~e^{\frac{\beta}{2} U(x)} \simeq p^{(i)}(x)~e^{\beta U(x)} \simeq \frac{1}{z_i} h^{(i)}(x)
\ee
In the hermitian formalism, the left- and right- state distributions are replaced by a single  distribution $\pi^{(i)}(x)$:
\be
\pi^{(i)} (x) \equiv    \frac{h_i(x)}{\sqrt{z_i}}~e^{-\frac{\beta}{2} U(x)} = \sqrt{z_i}~~\times~
\begin{cases}
 e^{-\frac{\beta}{2}U(x)} ~a^{(i)}(x) \\
 e^{+\frac{\beta}{2}U(x)}~p^{(i)}(x)
 \end{cases}.
\ee

In practical applications, MSMs are never defined by directly diagonalizing $H_{FP}$, as in this purely theoretical discussion. 
Instead, they  are built by analyzing an ensemble of  MD trajectories by means of dimensional reduction methods such as Time-lagged Independent Component Analysis (TICA) \cite{TICA1, TICA2, TICA3, TICA4} or Principal Component Analysis (PCA) \cite{TICA1, PCA} and by geometric clustering of the projected configurations. In this case, the cut-off scale $\Lambda$ is identified with the frequency of the slowest spectral component which is projected out, e.g. the inverse of the TICA timescale associated to the first excluded independent component.

Once the dynamics  has been coarse-grained at the level of microstates, the explicit dependence on the $3 N_a$-dimensional configuration point $x$ becomes redundant.  It is then convenient to introduce a formalism  in which such a dependence is  removed altogether. Hence, from this point on we shall use  the hermitian formulation of the Fokker-Planck dynamics (\ref{SE}) and adopt Dirac's ``bra-ket"  formalism, in which  the $N$ normalized microstates $\{|i\ra\}_{i=1,\ldots,N}$ are identified with points in a Hilbert space  $\mathcal{H}$. 

To construct the microstates $|i\ra$, we begin by introducing position eigenstates $|x\ra$ and the corresponding hermitian operator $\hat X$:
\be
\hat X |x \ra = x | x \ra  \;,
\ee
where the eigenvalues $x$ of $\hat X$ are points in the $3N_a$-dimensional  configuration space. Notice that the position eigenstates obey the normalization condition
\be
\la x| x' \ra = \delta(x-x').
\ee

The ket- and bra-microstates, $|i\ra $ and $\la i|$ are defined from the hermitian components of the state distributions $\pi^{(i)}(x)$:
\be\label{microket}
|i\ra &=& \int dx~ \pi^{(i)}(x) |x \ra\\
\label{microbra}
\la i| &=& \int dx ~\pi^{(i)}(x) \la x| \qquad (i=1, \ldots, N)
\ee
Note that they approximatively form an orthonormal set:
\be
\la i | j \ra  =  \int dx ~\pi^{(i)}(x)~\pi^{(j)}(x) = \delta_{ij} 
\ee
At finite temperature, orthogonality is weakly violated by the exponentially small overlaps between the microstates. 

The instantaneous configuration of a system is described by a time-dependent state $|\psi(t)\ra$, defined by 
\be\label{|v ra}
|\psi(t)\ra &=& \int dx ~\psi(x,t) |x\ra, 
\ee
where $\psi(x,t)$ is the solution of the imaginary-time Schr\"odinger equation (\ref{SE}).
In appendix \ref{app1}, we show that if the distribution $\psi(x,t)$ can be expressed through a  linear combination of eigenstates  below the gap -- like in Eq. (\ref{time-dep-decomp}) --, then the state $|\psi(t)\ra$  can be expressed as time-dependent linear combinations of the ket-microstates $|i\ra$:
\be\label{sumk}
| \psi(t) \ra &=& \sum_i \frac{n_i(t)}{\sqrt{z_i}}~|i \ra.
\ee
The $N$ time-dependent coefficients
\be
n_i(t) &=& \sqrt{z_i}~\la i | \psi(t) \ra \nn\\
&=&  \sqrt{z_i}~\int dx~ \pi^{(i)}(x) ~\psi(x,t) \nn\\
&=& \int dx~h_i(x)~P(x,t)
\ee 
are interpreted as the probabilities of observing the molecule in the different microstates at time $t$ and will be called the instantaneous  microstate populations. In  the low-temperature limit, the  time spent by the system in crossing high-energy regions is exponentially small compared to the time spent in the states. Consequently, the following sum-rule holds:
\be
\sum_{i=1}^N \sqrt{z_i} \la i | \psi(t) \ra = \sum_{i=1}^N~n_i(t)  \simeq 1.
\ee

Note that the normalization factor $\sqrt{z_i}$ ensures that the correct equilibrium populations are attained in the long-time limit:
\be
\lim_{t\to \infty} n_i(t) = \lim_{t\to \infty} \int dx~ P(x, t)~h_i(x) = \frac{1}{Z}~\int dx ~e^{-\beta U(x)} h_i(x) = \frac{z_i}{Z}.
\ee
where $Z=\sum_{i=1}^N z_i$ is the system partition function, in the low-temperature limit. 

The action of operators on the arbitrary  state $|\psi(t)\ra$ is  defined by its action on the microstates $|i\ra$, i.e.
\be\label{Odef}
\hat O ~|\psi(t) \ra \equiv \sum_{i j=1}^N ~O_{ij}~ | i \ra\la j| \psi(t) \ra = \sum_{i j=1}^N ~\frac{n_{j}(t)}{\sqrt{z_j}}~ O_{ij}~ | i \ra 
\ee 
where $O_{ij}\equiv \la i|\hat O |j\ra$ is a $N\times N$ matrix.

\subsection{Dynamics in the Space of Microstates}\label{sec_micros}

 To determine the dynamics of the state $|\psi(t) \ra$ it is convenient to introduce a time-evolution operator, $\hat U_\tau$
 \begin{equation}
 |\psi(t+\tau)\rangle = \hat U_\tau ~|\psi(t)\rangle
 \end{equation}
The ''wave-function'' $\psi(x,t)$ evolves according to 
\begin{equation}
\psi(x,t+\tau) = \la x| \psi (t+\tau) \ra = \int dx' \la x| \hat U_\tau |x'\ra \la x'|  \psi(t) \ra =  \int dx' \la x| \hat U_\tau |x'\ra~\psi(x', t).
\end{equation}
where $\la x| \hat U_\tau |x'\ra$ is  the imaginary-time propagator
  \begin{equation}
\la x| \hat U_\tau |x'\ra = \la x| e^{- \tau \hat H} |x'\ra
\end{equation}
and $\hat H$ will be called the (effective) Hamiltonian operator. 
 
We are now in a condition to derive  the time-evolution of the populations of the microstates $n_i(t)$: 
\be\label{dn}
 n_i(t+\tau) &=&  \sqrt{z_i} \la i | \psi(t+\tau)\ra =  \sum_{j=1}^{N} \sqrt{\frac{z_i}{z_j}}\la i|e^{-\hat H \tau}|j\ra~n_j(t) \nn\\
 \label{DTME}
 &\equiv&\sum_{j=1}^{N} ~T_{i j}(\tau)~ n_j(t). 
\ee
 The so-called \emph{transition probability matrix}
 \be\label{Tij}
 T_{ij}(\tau) = T_{|j\rangle \to |i\rangle}(\tau) \equiv  \sqrt{\frac{z_i}{z_j}} ~ \la i|e^{-\hat H  \tau}|j\ra
 \ee
expresses the probability that a system prepared in microstate $j$ is found in microstate $i$ after a  time interval $\tau$. In the MSM literature, the time interval  $\tau$ is ofter referred to as \emph{lag-time}. We emphasize that the matrix \eqref{Tij} manifestly satisfies the detailed balance condition, which implies that $T_{ij}(\tau)$ is a left stochastic matrix:
\begin{equation}
\sum_{i=1}^{N} T_{ij}(\tau) = 1 \quad \forall \tau \; \forall j = 1,\ldots, N.
\end{equation}

If the lag-time $\tau$ is finite,  Eq. \eqref{DTME} defines a so-called  discrete-time master equation.  Conversely,  if $\tau$ represents an infinitesimal time interval, Eq. \eqref{DTME} 
defines as a so-called continuous-time master equation (see e.g. \cite{Sriraman} for related discussions), 
\be\label{CTME}
\dot{n}_i(t) &= & \sum_{j=1}^{N} K_{ij} n_j(t) \qquad K_{ij} = 
\begin{cases}
k_{ij} \geq 0 &\text{ if } i \not = j \\
-\sum_{i} k_{ij} &\text{ if } i = j
\end{cases}
\ee

\begin{figure}[t!]
\begin{center}
\includegraphics[width=14cm]{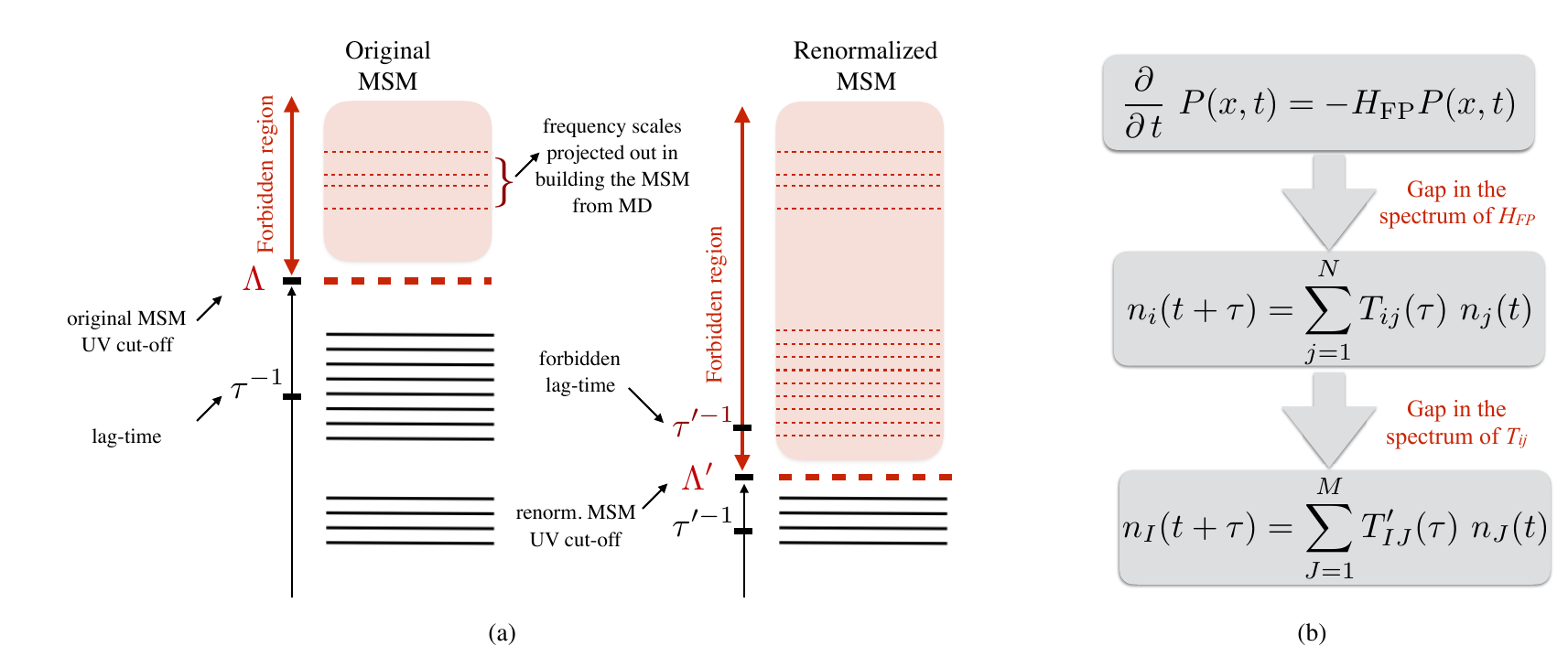}
\caption{a: The different frequencies scales must be ordered according to a specific hierarchy for the MSM to represent an ET of MD. The forbidden region is inaccessible to the ET. The right panel shows how the cut-off and forbidden region is modified by renormalisation (dimensional reduction). b: representation of the coarse-graining steps required to go from the microscopic Fokker-Planck dynamics to the dimensionally reduced MSM. }
\label{cutoffFig}
\end{center}
\end{figure}

The $K_{ij}$ is called the \emph{transition rate matrix} (sometimes also termed kinetic matrix or Markov generator). Its relationship with the transition probability matrix is simply given by
\begin{equation}
K_{ij} = \lim_{\tau \to 0} \frac{T_{ij}(\tau) - \delta_{ij}}{\tau} \; .
\end{equation}

We emphasize again that, in the RG perspective, the symbol $\frac{d}{dt}$ in Eq. (\ref{CTME}) does not represent a real derivative, but rather  a finite increment. Indeed, $dt$ cannot  be set smaller than the inverse frequency cut-off scale $\Lambda$. 
From this discussion it follows that the lag-time $\tau$ of discrete-time MSMs  must obey $\tau> dt\gtrsim 1/\Lambda$. Models which violate this inequality cannot be interpreted as rigorous ETs of MD. The hierarchies of different frequencies scales which must be obeyed for a MSM to represent an ET  is schematically illustrated in Fig.\ref{cutoffFig}.

Finally, the equation expressing the evolution of the states in the Hilbert space is
\begin{equation}\label{Hdef}
\begin{split}
-\frac{d}{dt} |\psi(t) \ra  &= \hat H | \psi(t) \ra \equiv \left[ - \sum_{ij} \sqrt{\frac{z_j}{z_i}} K_{ij} |i \rangle \langle j | \right] |\psi(t) \rangle \; .
\end{split}
\end{equation}
Notice that this equation provides the discrete formulation of the effective Schr\"odinger equation (SE) and that the symmetry of the following matrices (as introduced in \cite{Sriraman})
\be \label{symm1}
H_{ij} &\equiv& \la i | \hat H |j\ra = - \sqrt{\frac{z_j}{z_i}}~ K_{i j} \\
\label{symm2}
\Xi_{ij}(\tau) &\equiv&   \sqrt{\frac{z_j}{z_i}} T_{ij}(\tau)
\ee
expresses the detailed balance condition.
It is straightforward to show that $H_{ij}$ and $K_{ij}$ share the same spectrum, up to an overall sign. Indeed, 
\begin{equation}
\sum_{j}H_{ij}v_j =  -\sum_{j}\sqrt{\frac{z_j}{z_i}}K_{ij} v_j = \lambda v_i
\end{equation}
By defining $w_j\equiv \sqrt{z_j} v_j$, one has $\sum_j K_{ij} w_j = - \lambda w_j$. The same argument  applies also to the relationship between $T(\tau)$ and $\Xi(\tau)$. 

\section{Even coarser-grained ETs:  Dimensional Reduction of MSMs}
\label{RSRG}

\begin{figure}[t!]
\begin{center}
\includegraphics[width=10cm]{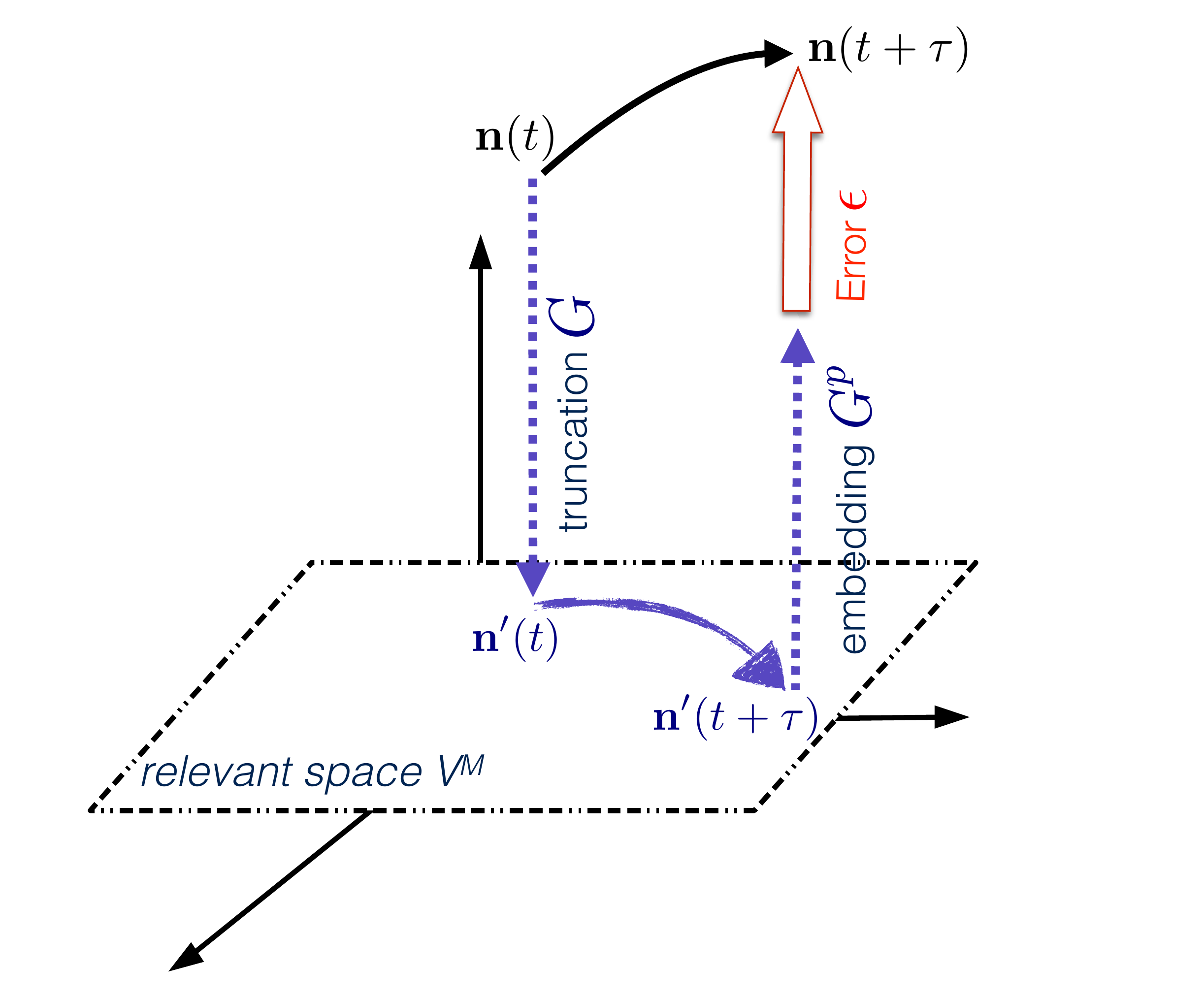}
\caption{Schematic representation of the dimensional reduction approach based on Real Space RG.}
\label{RSRGFig}
\end{center}
\end{figure}

In the previous sections we have seen how it is possible to derive a MSM which provides the rigorous ET for the system's slow relaxation dynamics. 
Now, we further assume that the spectrum of the associated master equation is also gapped --see Fig.\ref{cutoffFig}b--, enabling to  separate its slow and fast relaxation processes.
Under such circumstances it is possible to use RG theory to further coarse-grain the dynamics, i.e. to build a lower dimensional MSMs, in which the number of macrostates is set by the number of eigenvalues below the gap.

To introduce such a dimensional reduction, let us begin by considering the case  of a continuous-time MSM.The  dynamics of the populations is determined by the master equation (\ref{CTME}) which can be re-written in the following form:
\begin{equation}\label{master1}
n_i(t +dt) = \sum_{j=1}^{N}(\delta_{ij} + dt ~K_{ij})~n_j(t) 
\end{equation}
We stress once again that $dt$ plays the role  the UV cut-off of this ET, hence it must be chosen longer than the inverse UV cut-off frequency, yet shorter than \emph{all} relaxation times predicted by the rate matrix. 

Geometrically, the set of microstate populations $\{ n_i\}_{i=1,\ldots, N}$ can be identified as the coordinates of a unit-norm vector belonging to the positive quadrant of a  $N$-dimensional vector space $V^N$ -- see Fig. \ref{RSRGFig}. 
Let us now  introduce the so-called truncation matrix $G$, which maps vectors belonging to $V^N$ onto some insofar unspecified $M$-dimensional linear subspace $V^M$.  In particular, the time-dependent population vector ${\bf n} \equiv (n_1, n_1, \ldots, n_{N})$ is transformed by $G$ as follows:
\be
\label{truncation}
n_i \mapsto n'_I \equiv \sum_{j=1}^{N}~G_{Ij}~n_j,\qquad
\left(\begin{array}{c}
 I=1, \ldots, M.\\
j=1,\ldots, N
 \end{array}\right)
\ee

From here on, we shall always denote with uppercase letters all the indices ranging from $1$ to $M$ and with  lowercase letters those ones ranging  from $1$ to $N$. Since $N>M$, the truncation matrix is obviously not invertible. However, it is possible to define its so-called pseudo-inverse matrix $G^p_{i J}$ which coincides with $G^{-1}$ for $M=N$ and satisfies the so-called Moore-Penrose relationships (see Appendix \ref{app3}). In practice, the pseudo-inverse of the truncation matrix (which we shall refer to as the embedding matrix) can be explicitly constructed from the singular value decomposition of the truncation matrix $G$:
\begin{equation}\label{psinv}
G_{Ij} = (U \Sigma V^T)_{Ij}\quad \Rightarrow \quad G^P_{iJ} \equiv ( V \Sigma^{-1}U^T)_{iJ},
\end{equation}
where $U_{IK}= u_I^{(K)}$ and $V_{ik}=v^{(k)}_i$ are the orthogonal matrices of the left- and right- singular vectors of $G$ and $\Sigma_{Kh}= \lambda_K ~\delta_{Kh}$ is the rectangular matrix with the singular values on the diagonal.

Using the embedding and truncation matrices, we can define a new lower-dimensional master equation:
\be\label{master2}
 n'_I(t+dt)  = \sum_{J=1}^{M} (\delta_{IJ}+ dt~ K'_{IJ}) ~n'_J (t) 
\ee
 where
\be\label{def_K'}
K'_{IJ} = ( GK G^P)_{IJ}
\ee
is an effective  $M\times M$ transition rate matrix.

Our goal is to  identify the  truncation and embedding matrices $G$ and $G^P$ for which the lower-dimensional master equation (\ref{master2})  provides the best possible Markovian low-dimensional approximation of the dynamics defined by the original $N-$dimensional master equation (\ref{master1}).  In this case, the components of the $M-$dimensional vector ${\bf n}'(t)$ are interpreted as
the probabilities of observing the system in each of the different macrostates and the matrix $K'$ contains the  interconversion rates between macrostates. 

Our general strategy is schematically represented in Fig. \ref{RSRGFig}:  we first introduce a norm which quantifies the difference between the dynamics described by the original master equation (\ref{master1}) and the reduced master equation (\ref{master2}), thus defines the error introduced  by the dimensional reduction. Next, we vary the choice of  the subspace $V^M$ until such a difference is reduced to a minimum.  The  vector space $\bar V^M$  which corresponds to the least difference will be called the \emph{relevant subspace}. 

To implement this scheme in practice, we begin by  embedding the effective dynamics given by Eq. (\ref{master2})  in the larger vector space $V^N$, where the original dynamics given by (\ref{CTME})  is defined.  This can be done by applying the embedding  matrix $G^P$ to Eq. (\ref{master2}):
\be\label{RGT}
n_i(t+dt)  =\sum_{j=1}^{N} \left( G^P (\Id + dt K') G \right)_{i j} ~n_j(t) \
\label{RGT2} = \sum_j R_{ij} n_j(t) +dt \sum_{j=1}^N \left(R K R\right)_{i j} ~n_j(t) \;.
\ee
where the projector
\be\label{proj_mat}
R_{ij} \equiv (G^P  G )_{ij}  = \sum_{K=1}^{M} v^{(K)}_i v^{(K)}_j
\ee
is called the reduction matrix. In this equation, $v_i^{(K)}$ is the $i-$th component  of the $K$-th right-singular vector of the truncation matrix $G$. 

It is now convenient to switch to the quantum-mechanical notation introduced in the previous section. We define the reduction operator $\hat R$ acting on the Hilbert space: 
\be\label{projector}
\hat R =  \sum_{K=1}^{M}|{\bf v}_K \ra \la {\bf v}_K | = \hat{\Id} - \sum_{k=M+1}^{N} |{\bf v}_k \ra \la {\bf v}_k |.
\ee
where the states $|{\bf v}_k\ra$ are called the target states and are related to the components of the right singular vectors of the truncation matrix $G_{Ij}$, i.e.
\be
v^{(k)}_i \equiv \la i | {\bf v}_k\ra.
\ee
Notice that, with this definition, the reduction operator $\hat R$ and the reduction matrix $R_{ij}$ are related by $\la i | \hat R | j\ra = R_{ij}$. Thus, finding the optimal dimensional  reduction is equivalent to  identifying  the  set of target states $\{|{\bf v}_k\ra\}_{K=M+1,\ldots N}$ whose elimination from the Hilbert space minimally affects the time evolution of the probability state $|\psi(t)\ra$. 

It is most convenient to perform this elimination by projecting out  one target state at the time. Let us therefore discuss the elimination of the first one, $|{\bf v}\ra$:
\be \label{R}
\hat R = \hat{\Id} - |{\bf v}\ra \la {\bf v}| = \hat{\Id} - \hat{\mathcal{P}}_{|\textbf{v}\rangle}
\ee
The difference between the evolution of the population vectors in the original and in the effective dynamics is written component-wise as follows
\be
n_i^{\text{original}}(t) - n_i^{\text{effective}}(t) &=& \sqrt{z_i}\langle i  |\psi(t+dt) \rangle_{\text{original}} - \sqrt{z_i}\langle i | \psi(t+dt) \rangle_{\text{effective}} \nn\\
&=& \sqrt{z_i}\langle i | \left[ \left( \hat{\Id} - dt ~\hat H \right) - \left( \hat{R} - dt ~\hat{R}~\hat H~\hat{R}\right) \right] |\psi(t)\rangle
\ee
Hence, the term into square bracket on the right-hand side expresses the error which is introduced in the dynamics by projecting out the target state $|{\bf v}\rangle$, thus is called the error operator. 
Recalling definition \eqref{R}, it can be written as follows:
\begin{equation}
\hat \varepsilon = \hat{\mathcal{P}}_{|\textbf{v}\rangle}  - dt~ \left( \left\{\hat H, \hat{\mathcal{P}}_{|\textbf{v}\rangle}  \right\}   - \la \textbf{v}|  \hat{H} | \textbf{v}\ra \hat{\mathcal{P}}_{|\textbf{v}\rangle} \right),
\end{equation}
where $\{ \cdot, \cdot \}$ denotes the anti-commutator. Our goal is then to minimize the Frobenius norm\footnote{There are many possible norms between which one can choose, \emph{a priori}. We follow the choice of Ref.s \cite{RG1, RG2}.} $|\hat{\varepsilon}|^2_F$ of the error operator with respect to the choice of target state $|{\bf v}\rangle$:
\begin{equation}
| \hat \varepsilon|^2_F \equiv \sum_{ij} |\langle i |\hat{\varepsilon} |j\rangle |^2 = \sum_{ij} \left[ v_{i} v_j - dt~ \left( \sum_l H_{i l} v_l v_j+ \sum_l v_i v_l~H_{lj} - \la {\bf v}|\hat{H}|{\bf v} \ra~v_i v_j~\right)  \right]^2.
\end{equation}
Expanding the square and retaining only the leading-order terms in $dt$ we find:
\be\label{eF}
| \hat \varepsilon|^2_F &=&1 - 2 dt \sum_{ij} v_i v_j H_{ij} + \mathcal{O}(dt^2)
\ee

The target vector components $v_i$  are found by extremizing the bilinear $\sum_{ij} v_i v_j H_{ij}$ under the normalisation  constraint   $\sum_i v_i^2=1$. 
Introducing a Lagrange multiplier $\lambda$ and imposing stationarity with respect to variation of $v_i$ we obtain:
\be\label{const_ot}
0 = \frac{\partial}{\partial v_k} \left[\sum_{ij} v_i v_j H_{ij} + \lambda \left(1-\sum_iv_i^2\right) \right] 
\ee
which yelds
\be
H_{ij} v_j= \lambda v_i = |\lambda|~v_i
\ee
Thus, we have found that all $N$ eigenvectors  of the orthogonal matrix $H_{ij}$ locally minimize  the target function inside the square brackets of  Eq. (\ref{const_ot}). In particular, if $v_i^{(k)} = \langle i | {\bf v}_k\rangle$ are the components of the  $k-$th eigenvector,
$
H_{ij} v^{(k)}_j= |\lambda_k| v^{(k)}_i,
$
then $\la {\bf v}_k| \hat{H}|{\bf v}_k \ra= \left|\lambda_k\right|$ and the error introduced by projecting out this vector is
\be
|\hat{\epsilon}|_F^2\simeq 1- 2 dt |\lambda_k|.       
\ee
The global minimum of our error estimator is realized by projecting out the eigenvector  of $H_{ij}$ with the largest eigenvalue. This procedure should be repeated to project out all eigenstates with eigenvalues above the gap (i.e. all system's eigenmodes with fast relaxation frequencies).  Indeed, from Eq. (\ref{eF})  it follows that the elimination of all these target vectors generates errors which are small and comparable. The elimination of all these states results in the lowering of the cut-off. The new UV scale  $\Lambda'$ is set by the frequency of the slowest mode which was projected out --see Fig. \ref{cutoffFig}a--.

The  renormalization of a discrete-time MSMs is completely analog  and is  reported in appendix \ref{app6}. 
We recall that, in the discrete-time formulation, one needs to specify the lag-time $\tau$. For the dimensionally reduced MSM to represent a rigorous ET, such a lag-time must be chosen in such a way to remain larger than the new UV time cut-off scale $1/\Lambda'$. In other words, the RG transformation of a discrete-time MSMs is intrinsically consistent only  if the lag-time of the original MSM is much longer than the inverse of the smallest relaxation frequency projected out during the dimensional reduction --see Fig.\ref{cutoffFig}a--.

\subsection{Identifying the Macrostates and Computing the Reduced Kinetic Matrix} \label{macro_section}
So far we have developed a procedure to  identify subspace of the original vector space $V^N$ where the relevant dynamics takes place. Let us now address the problem of identifying the macrostates.
 We recall that we are  working under the hypothesis that also the spectrum of the  effective hamiltonian operator $\hat H$ is gapped, namely that the first $M$ right eigenstates
\be
\hat H | {\bf v}_K\ra = \lambda_K |{\bf v}_K\ra
\ee
have eigenvalues   $\lambda_1, \ldots, \lambda_M \sim \lambda$ well separated by all other eigenvalues $\lambda_{M+1}, \ldots, \lambda_N \gtrsim \Lambda$, with $\Lambda \gg \lambda$. In the previous section, we  discussed how the existence of a gap in the spectrum of the Fokker-Planck operator leads to the definition of $N$ continuous state distributions $p^{(i)}(x)$, associated to the system's microstates. The same arguments can be repeated at the discrete level and lead to the definition of macrostates. We expect to to find  $M$ linear combinations of the lowest $M$ right-eigenvectors of  the hermitian Hamiltonian operator $\hat H$, 
\be \label{macrostate}
|{J}\ra \equiv \sum_{K=1}^M T_{J K} |{\bf v}_K\ra \quad J = 1, \ldots, M,
\ee
which  obey the following properties:
\begin{enumerate}
\item The coefficients $\Pi_i^{(J)}$  expressing the ket-macrostates $|J\rangle$ as linear combination of the microstates, 
\be\label{lin}
|J \rangle = \sum_{i=1}^N \Pi_i^{(J)} |i\rangle 
\ee
are non-negative, $\Pi_i^{(J)}\ge 0$. Those coefficients are the analog of the $\pi^{(i)}(x)$ entering Eq. (\ref{local_gibbs}). 
\item The $M$ states $\{|J\rangle\}_{J=1,\ldots,M}$  are  disjoint, that is  $\Pi_i^{(J)} \Pi_i^{(K)} \sim O\left( \frac{\lambda}{\Lambda} \right) $, for all $K\ne J$.  Disjointness implies that  each microstate $|i\rangle$ belongs to one and only one of the $|J\ra$ states.
\end{enumerate}
One can immediately notice that Eq. \eqref{macrostate} guarantees that the projector operates as the identity in the relevant space $V^M$:
\begin{equation}
\hat{R}|J\rangle = \sum_{H=1}^M |{\bf v}_H\rangle \langle {\bf v}_H| \sum_{K=1}^M T_{JK} |{\bf v}_K\rangle = \sum_{K=1}^M T_{JH} |{\bf v}_H\rangle = |J\rangle
\end{equation}
This means that the projector operator $\hat{R}$ can be rewritten as a projector on the macrostates:
\begin{equation}\label{new_proj}
\hat{R} = \sum_{J=1}^M |J \rangle \langle J|.
\end{equation}
Given Eq. \eqref{new_proj}, let us compute the projector matrix elements on the microstates basis:
\begin{equation}
(G^PG)_{ij} = R_{ij} = \langle i | \hat{R} | j \rangle = \sum_{J=1}^M \langle i |J \rangle \langle J| j \rangle = \sum_{J=1}^M \Pi_i^{(J)} \langle J | j \rangle
\end{equation}
From this we deduce that 
\begin{equation}\label{GP}
G_{iJ}^P = \Pi_{i}^{(J)}.
\end{equation}
Furthermore, to make sure that the relation $R_{ij} = (G^PG)_{ij}$ is satisfied, one necessarily has to define the bra-macrostates $\langle J |$ as
\begin{equation}\label{G}
\langle J | = \sum_{i=1}^N \left(\Pi_{i}^{(J)}\right)^P \langle i | = \sum_{i=1}^N G_{Ji} \langle i |.
\end{equation}
Thus, the expansion coefficients of the macrostates in the microstates basis naturally provide the truncation and embedding matrices. We recall that the condition of minimum Frobenius norm of the error operator poses a constraint only on the reduction matrix $R$, thus leaving some arbitrariness on the choice of $G^P$ and $G$. Within the manifold of  $G^P$ and $G$ matrices which satisfy $R=G^P G$, only  the choice given by (\ref{GP}) and (\ref{G}) leads to the correct probabilistic interpretation, i.e. ensues that  $G_{Ji}\ge 0$ is related to the probability of observing the microstate $i$ inside the macrostate $J$. 

Let us finally tackle the problem of how constructing  the macrostates starting from the set of microstates of the system. To this goal, we  consider a new set of $N$ states defined by:
\be
|M_i\ra = \hat{R}|i\rangle \qquad \langle M_i| = \langle i |\hat{R}
\ee
Then,
\begin{equation}\label{M_states}
\begin{split}
\langle j | M_i \rangle &= \langle j | \hat{R} | i \rangle = R_{ij} \\
\langle M_i | j \rangle &= \langle i | \hat{ R} | j \rangle = R_{ji} = R_{ij}.
\end{split}
\end{equation}
where the second line in Eq. \eqref{M_states} follows from the fact that $\hat R$ is a projector, so $\hat R^\dagger = \hat R$. Let us now suppose that some microstate $| i \ra$ is contained in the macrostate $|I\ra$ (i.e. $\Pi _i^{(I)} > 0$). Then, it is immediate to prove that the state $|M_i\ra$ state has non-zero overlap with $|I\ra$:
\begin{equation}
\la I |M_i\ra = \la I|\hat R |i \ra = \la  I|i \ra = \Pi_j^{(I)} > 0
\end{equation}
where we used the fact that $|M_i\ra$ belongs to the relevant subspace. Conversely, if the microstate $|i\ra$ is not contained in the macrostate  $|I\ra$, then $|M_{i}\ra$ and $|I\ra$ have no overlap (up to correction of order $O\left(\frac{\lambda}{\Lambda}\right)$):
\be
\langle I | M_i\ra = \la I | i \ra \sim O\left(\frac{\lambda}{\Lambda}\right).
\ee
Hence, if  the state $|M_{i}\ra$ overlaps with $|I\ra$, then it has no relevant overlap with any other macrostate $|J\ra$.  This implies that each of the $|M_1\ra, \ldots |M_N\rangle $ states is  either null or parallel to one and only one macrostate.

Given this consideration, we can compute the coefficients $\Pi_i^{(I)}$ of the generic macrostate $|I\rangle$ by imposing the same normalization adopted for the macrostates:
\begin{equation}
\delta_{ij} = \langle M_i| M_j\rangle = \langle i | \hat{R} \hat{R} | j \rangle = \langle i | \hat{R} |j \rangle = R_{ij}
\end{equation}
This means that the coefficients are simply provided by
\begin{equation}\label{GPi}
\Pi_i^{(I)} = G^P_{iI} = \frac{R_{iI}}{\sqrt{R_{ii}}}
\end{equation}
where $R_{Ii}$ are the $M$ linearly independent rows corresponding to the arbitrary choice of the $M$ linearly independent $|M_i\rangle$ states. In Appendix \ref{app5} we show that $R^P_{iI} = R^T_{iI}$, which implies
\begin{equation}\label{G_norm}
G_{Ii} = \frac{R_{Ii}^T}{\sqrt{R_{ii}}}.
\end{equation}

The last ingredient of our Renormalization Group procedure concerns computing the effective kinetic matrix. We start from the effective Hamiltonian operator represented in the macrostates basis
\begin{equation}\label{K_first}
H'_{IJ} = \langle I | \hat{H} | J \rangle = \sqrt{\frac{Z_J}{Z_I}} K'_{IJ}
\end{equation}
where $Z_I$ is the partition function of the $I-$th macro state 
On the other hand, 
\begin{equation}\label{K_first_}
H'_{IJ} = \langle I | \hat{H} | J \rangle = \sum_{i,j=1}^N \langle I | i \rangle \langle i | \hat{H} | j \rangle \langle j | J \rangle = \sum_{i,j=1}^N G_{Ii} H_{ij} G_{jJ}^T
\end{equation}
Plugging Eq. \eqref{K_first} into Eq. \eqref{K_first_} one finds that the effective kinetic matrix is the solution of the following equation
\begin{equation}\label{K'}
K'_{IJ} = \sqrt{\frac{Z_I}{Z_J}} \sum_{i,j=1}^N G_{Ii} H_{ij} G_{jJ}^T.
\end{equation}
We emphasize that this is an implicit relationship. Indeed the $M$ partition functions $Z_I$ in  the right-hand-side are the components of the lowest right-eigenvalues of the effective kinetic matrix $K'_{IJ}$. 

All the proofs given in this section hold also for discrete-time MSMs, since $\hat{H}$ and $e^{-\hat{H}\tau}$ share the same eigenvectors $\forall \tau$. Thus, the macrostates of a system described by a transition probability matrix $T_{ij}(\tau)$ are simply provided by Eqs. \eqref{GP} and \eqref{G_norm} and the corresponding effective transition probability matrix is readily obtained as
\begin{equation}\label{T'}
T'_{IJ}(\tau) = \sqrt{\frac{Z_I}{Z_J}} \sum_{i,j=1}^N G_{Ii} \Xi_{ij}(\tau) G_{jJ}^T
\end{equation}
where $\Xi_{ij}(\tau)$ is defined in Eq. (\ref{symm2}).

To summarize, Eq.s (\ref{GPi}), (\ref{K'}) and (\ref{T'}) are  the most important results of this paper. Indeed, Eq. (\ref{GP}) provides the definition of macrostates in terms of microstates, while Eq.s (\ref{K'}) and (\ref{T'}) provide the  new continuous- and discrete-time master equations, which approximate the MD  in the reduced space of macrostates. 

\subsection{The Renormalization Group Clustering Algorithm}\label{RGC}

We now describe our Renormalization Group Clustering (RGC) algorithm,  which implements the RG theory developed so far.  Here we illustrate it for a continuous-time MSM. The formulation of the same algorithm for a discrete-time MSM is analog and is reported in Appendix \ref{app6}.

\begin{enumerate}
	\item Compute the spectrum of relaxation frequencies $ \{\lambda_k\}_k$ from the kinetic matrix $K$. Retain only the frequencies below the gap, i.e. up to $k=M$, where $\left| \lambda_M \right| \ll \left| \lambda_{M+1} \right|$, and the corresponding set of  right eigenvectors. Let us call $Z$ the eigenvector corresponding to $\lambda_1=0$ and define
	\begin{equation}\label{delta_cont}
	\Delta = \left| \frac{\lambda_M}{\lambda_{M+1}} \right|\ll 1;
	\end{equation}
	which quantifies the extent the spectral gap. 
	\item Compute the hamiltonian matrix $H_{ij} = \sqrt{\frac{z_j}{z_i}}K_{ij}$;
	\item Build the projector $R_{ij} = \sum_{i=1}^M v_i^{(K)}v_j^{(K)}$ where $v_i^{(K)}$ is the $i$-th components of the $K$-th lowest eigenstate of the hamiltonian matrix;
	\item Extract the $M$ linearly independent vectors from $R$ to build the matrix $G$;
	\item The matrix $G$ may contain entries of order $\mathcal{O}\left(\Delta\right)$, some of which may be negative. To preserve only the relevant terms and the probabilistic interpretation,  filter out all such  terms using the following rule: A matrix element $G_{Ij}$ is set to $0$ if there exists a $G_{Ik}$ element such that
	\begin{equation}\label{spurious}
	\left| \frac{G_{Ij}}{G_{Ik}} \right| \leq \Delta.
	\end{equation}
	Once the irrelevant terms have been deleted from the matrix, normalize its rows dividing them by $\sqrt{R_{ii}}$ (which is equivalent to normalizing to 1 the sum of the elements in each of the rows of the $G$ matrix);
	\item Solve the self-consistent equation $K'_{IJ} = \sqrt{\frac{Z_I}{Z_J}} \sum_{i,j} G_{I i} H_{ij} G^T_{jJ}$ to obtain the reduced kinetic matrix. To this goal, one may use the fixed-point algorithm described Appendix \ref{app7};
	\item Compute the spectrum of $K'$ and isolate the smallest mode $\lambda'_1$. In general, $\lambda'_1 \propto \Delta$. To ensure the existence of a stationary state, redefine the $K'$ spectrum  as
	\begin{equation}\label{shift}
	\lambda'_K\to  \lambda_K' - \lambda'_1
	\end{equation}
	and recompute the effective kinetic matrix as
	\begin{equation}
	K' = V \text{diag}[\lambda'_1, \lambda'_2, \ldots, \lambda'_M] V^{-1}
	\end{equation}
	where $V$ is the orthogonal matrix which diagonalizes $K'$.
\end{enumerate}
A Python code implementing this algorithm and the equivalent version for discrete-time MSMs can be made available by the authors upon request.

We emphasize that the effective transition rate matrix $K'$ (or, equivalently, the transition probability matrix $T'$) is expected to satisfy the microscopic reversibility condition only up to corrections of $\mathcal{O}\left(\Delta\right)$.  On the other hand, to solve the self-consistent equation and to determine the equilibrium population it is important to consider  effective models which obey such a condition. In the RGC algorithm this is done at step 7, by shifting the spectrum of the effective matrix by a factor $\lambda'_1$ so that the lowest eigenvalue of $K'$ is null. 
We also stress the fact that the results of the renormalized MSM are only expected to be accurate up relative corrections which are expected to scale according to the gap ratio $\Delta$. 

\section{Illustrative Examples}
\label{tests}

In this section we provide two simple illustrative applications of the RGC algorithm based on simple toy models.

\subsection{MSM with Multiple Gaps in the Spectrum of Relaxation Frequencies}
\label{toy1}
\begin{figure}[!h]
\includegraphics[width=14cm]{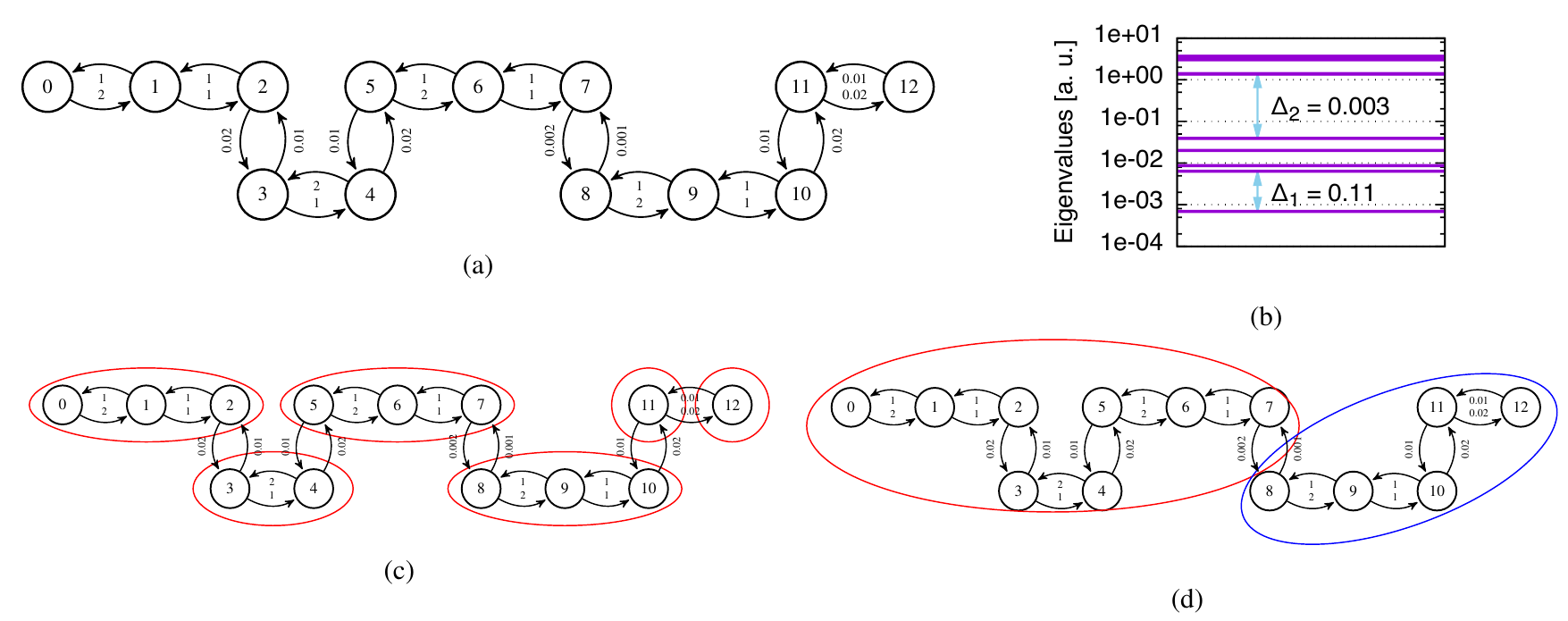}
    \caption{(a) Graphical representation of the 13 microstates of the MSM analyzed in \ref{toy1}; (b) spectrum of the corresponding kinetic matrix. Eigenvalues are expressed in units of $k$. The zero-mode is not shown; (c) graphical representation of the six macrostates obtained by clustering with respect to the gap $\Delta_2$. (d) graphical representation of the two macrostates obtained by clustering with respect to the gap $\Delta_1$.}
\label{fig1}
\end{figure}

As a first example of application of the RGC algorithm, we consider the MSM represented in Fig. 4a, composed of $13$ microstates. The arrows represent the connectivity between the different states and the numbers near the arrows express the corresponding transition rates (measured in units of some reference rate $k$).  In particular, we notice that most transition rates are of order $k$, a few of them are of order $k \times 10^{-2}$ and two are of order $k \times 10^{-3}$.
Physically, this model represents a system in which the relaxation to thermal equilibrium is slowed down by two energy barriers playing the role of kinetic bottlenecks. 
The $K_{ij}$ and $H_{ij}$  matrix are respectively showed in Eq. \eqref{app_K} and \eqref{app_S} of Appendix \ref{matrices}.

Fig.4b shows that the spectrum of the matrix $K_{ij}$ displays two well  separated gaps,  denoted with $\Delta_1$ and $\Delta_2$, respectively. 
Eigenvalues above $\Delta_2$ are of order $k$,  those between $\Delta_1$ and $\Delta_2$ are of order $k \times 10^{-2}$, while the two below $\Delta_1$ are respectively $0$ and $\sim k \times 10^{-3}$. Consequently, the original continuous time MSM must be defined using a cut-off $dt$ smaller than the inverse of the highest frequency shown in Fig.~4b. 

In the presence of two gaps it is possible to define two ETs, characterized by a different level of resolution, i.e. a different degree of dimensional reduction. In the finer-grained effective MSM, only the fast relaxation modes above the $\Delta_2$ gap are projected out. The number of macrostates of this effective model is set by the number of eigenvalues below the $\Delta_2$ gap, and the new cut-off $\Lambda'$ is smaller than the smallest relaxation frequency above the $\Delta_2$ gap. 
In the even coarser-grained effective MSM, all the relaxation modes with frequency above the $\Delta_1$ gap are projected out and a new cut-off $\Lambda''< \Lambda'$ is defined correspondingly. In this ET, there are only two macrostates.
 
From the block diagonal structure of the kinetic matrix in Eq. (\ref{app_K}) it is immediate to guess the structure of the macrostates in the finer- and coarser-grained effective MSM, which are graphically highlighted by the ellipses in Fig.4c and Fig.4d, respectively.
In the following we show how these structures emerge from the RGC analysis. 

The first step  consists in computing the reduction matrix  $R_{ij}$, i.e.  the projector onto the lowest 6 eigenstates of the $H_{ij}$ matrix. The result is shown in Eq. \eqref{app_R} of Appendix \ref{matrices}.
We note that some rows of this matrix differ only by relative corrections of order $10^{-3}$, i.e. by an amount of the order $\Delta_2$, which represents the estimate of the  intrinsic relative uncertainty of the RGC approach. 
To this level of accuracy, it is legittimate to set to zero all entries of relative order $10^{-3}$ and require that two vectors are equal if their relative difference is lower than $10^{-3}$. The new matrix is reported in Eq. \eqref{app_R'} of Appendix \ref{matrices}. It is worth noticing that such filtering procedure preserves the symmetry of the original projector. Furthermore, in this form it is clear that only 6 lines are linearly independent. The truncation operator is obtained by normalizing and arranging in raw such lines and is provided in Eq. \eqref{app_G} in Appendix \ref{matrices}.

Finally, the corresponding effective kinetic matrix is calculated by solving the self-consistent equation (\ref{K'}). The result obtained using the fixed-point algorithm described in Appendix \ref{app7} is reported in Eq. \eqref{app_K'1} of Appendix \ref{matrices}.
We note that that the raws of this matrix do not sum exactly to $0$, but to a number of the order $10^{-5}$, smaller than the intrinsic error of the method. Within such an accuracy it is legitimate to shift the diagonal element by an amount given by the lowest eigenvalue of the matrix (in this particular case $\sim 2\times 10^{-5}$) so that the new $K'$ satisfies detailed balance condition. This spectral shift procedure yields the matrix reported in Eq. \eqref{app_K'2} in Appendix \ref{matrices}.

The same procedure can be repeated to build the coarser grained effective MSM. In this case, we need to build a new projector matrix $R$ which maps onto the space spanned by the 2 lowest eigenvectors of $H_{ij}$. To build the truncation operator we discard all relative differences which are of order $10^{-2}$ or lower and obtain the truncation operator showed in Eq. \eqref{app_G2} in Appendix \ref{matrices}.  Through this procedure one finds the two macrostates showed in Fig.4d. The kinetic matrix obtained after enforcing detailed balance is reported in \eqref{app_K'3} in Appendix \ref{matrices}.

Let us now assess the accuracy of these two effective descriptions in describing the kinetics. To this goal, we compare the  relaxation rates $\lambda_k$ of the two effective models with the slowest rates calculated from the eigenvalues of the kinetic matrix of the original MSM.
We expect the finer-grained MSM to reproduce well the lowest 6 relaxation frequencies of the original model, while the coarser grained model to reproduce only the thermal relaxation rate (we recall that first eigenvalue is set to 0 by construction).

\begin{table}[!h]
\begin{tabular}{c|cccccc}
\hline \hline
 &  $\lambda_R$ & $\lambda_3$ & $\lambda_4$ & $\lambda_5$ & $\lambda_6$ \\ \hline
Original relaxation rates & 6.93$\times 10^{-4}$ & 6.39$\times 10^{-3}$ & 8.74 $\times 10^{-3}$ & 2.024 $\times 10^{-2}$ & 3.9328 $\times 10^{-2}$ \\ 
Effective relaxation rates (6 macrostates) & 6.88 $\times 10^{-4}$ & 6.41 $\times 10^{-3}$ & 8.72 $\times 10^{-3}$ & 2.028 $\times 10^{-2}$ &   3.9323 $\times 10^{-2}$ \\
Relative error & 0.7$\%$ 	& 0.3$\%$	 & 0.2$\%$ & 0.2$\%$ & 0.01$\%$ \\ 
Effective relaxation rates (2 macrostates) & 7.04$\times 10^{-4}$ & - & - & - & - \\ 
Relative error & 1.6$\%$ 	& - & - & - & - \\ 
\hline
\end{tabular}
\captionof{table}{Comparisono between the relaxation rates of the original 13 microstates system and the effective relaxation rates obtained, respectively, from the six and two macrostates models. Relative errors of the latter ones with respect to the original relaxation rates are also shown.}
\label{tab:timescales1}
\end{table}

 The results are presented in Table  \ref{tab:timescales1}.  We see that the  predictions of the effective theories  are always very close with those of the  original theory. The difference between the relaxation frequencies calculated in the original and in the effective models range from 10$^{-3}$ to 0.2 \%. 
We note that the errors of the coarser model are about one order of magnitude bigger than the those of the finer model, reflecting the fact that the gap in the coarser model  is smaller. 

\subsection{Emerging Two-State Kinetics in a Funnelled Energy Landscape}\label{toy2}
\begin{figure}[t!]
\begin{center}
\includegraphics[width=14cm]{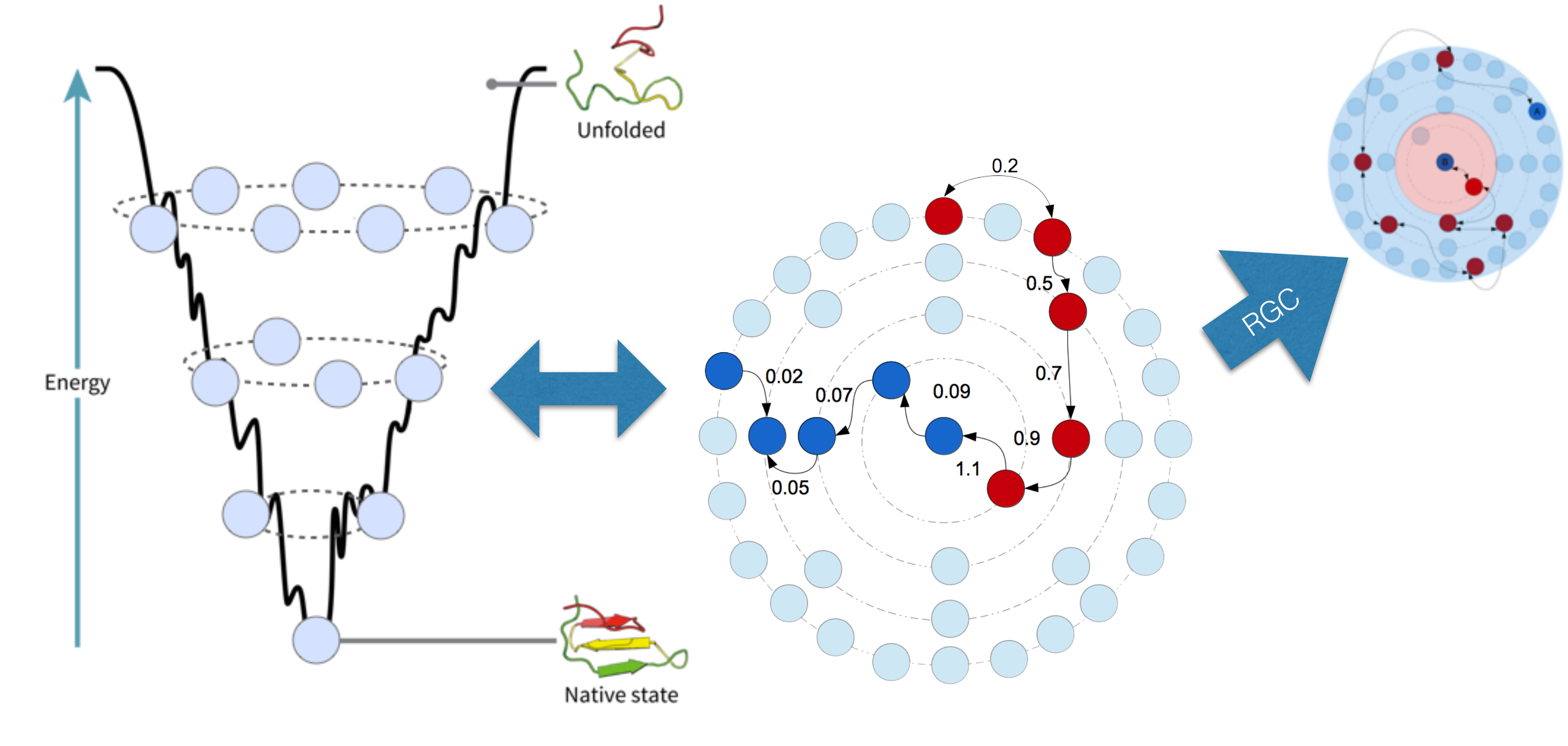}
\caption{Graphical representation of a funnel energy landscape and the corresponding MSM used in \ref{toy2}. The beads are interpreted as the microstates of the system and the arrows highlight a particular folding  pathway. 
The figure in the upper-right corner depicts the two macrostates of the system, identified with the folded and unfolded states, generated by the RGC algorithm.}
\label{funnel_sys}
\end{center}
\end{figure}

An important question to address is whether the RGC algorithm remains reliable also for entropy-dominated free-energy barriers.
To answer, we consider a MSM which mimics protein folding kinetics and is inspired by the so-called Energy Landscape Theory \cite{EL}.

In ELT, the energy surface of proteins is pictured as a relatively smooth funnel, with a single global minimum, corresponding to the protein native state (see Fig. \ref{funnel_sys}). The energy function  decreases with increasing structural overlap between the instantaneous chain configuration and the  native configuration. 
On the other hand, the conformational  entropy is assumed  decrease more rapidly, with increasing overlap. The resulting energy-entropy competition can give rise to a single, entropy-dominated  free-energy  barrier which separates the so-called unfolded state from the native state. In this case, the folding kinetics can be effectively described by  a single relaxation rate.
 
The cartoon in Fig. \ref{funnel_sys} represents our MSM, which was engineered in order to reproduce these features. The  microstates are arranged in concentric rings, with the  states belonging the outer rings representing  configurations with smaller structural overlap with the native conformation.  
All microstates belonging to a given ring are kinetically connected to all others in the same ring and to all states belonging to the neighbouring inner and outer rings. 

The energetic bias towards the native state (center of the rings)  is introduced by promoting the matrix elements of the kinetic matrix associated to inwards  transitions. 
The entropic effect  favouring denaturation is obtained by imposing that the number of states belonging to a ring  rapidly increases with the radius of the ring. In particular, here we discuss a MSM model consisting of  5 concentric rings, containing 1000, 200, 50, 5, 1 states, respectively. The transition rates which have been used to define the kinetic matrix are showed in Fig. \ref{funnel_sys}. 

By applying the RGC algorithm described in the previous section and illustrated in the previous example, one obtains a drastic simplification of the kinetics of this model. Indeed, the resulting effective MSM contains only two macrostates, which can be identified with the folded and unfolded states. The corresponding renormalized master equation is then 
\begin{equation}
\begin{cases}
\frac{d n_F}{dt} = k_f n_U - k_u n_F\\
\frac{d n_U}{dt} = k_u n_F - k_f n_U \end{cases}
\end{equation}
where $n_F$ is the population fraction in the folded state, $n_U$ is the population fraction in the unfolded state,  while $k_f$ and $k_u$ are respectively the folding and unfolding rates.  The macrostates after renormalization are represented  as the shaded areas in upper-right corner of Fig. \ref{funnel_sys}. We note that the folded state contains a few microstates, so in this specific MSM,  the native dynamics is not trivial. A similar feature of native dynamics will also appear in the model discussed in section {MDtest}, in which the MSM was obtained directly by reducing an ultra-long MD trajectory. 
In Tab. \ref{tab:fun_kin_rates} we report the relaxation timescales and transition rates obtained in the original model and after applying the RGC algorithm. 
We see that, also in the presence of an entropy driven barrier, the relaxation kinetics is reproduced to a very high degree of accuracy by the effective model.  

\begin{table}[!h]
\centering
\begin{tabular}{ccccccc}
\hline \hline
$\lambda_R$ & $\lambda'_R$ & $\delta \lambda'_R$ & $k_f$ & $k_u$ & $\delta k$ \\ 
\hline
5.4960 & 5.4957 & 5$\times 10^{-3}\%$ & 5.4957 & 2 $\times 10^{-7}$ & 2$\times 10^{-4}$\\ 
\hline
\end{tabular}
\caption{Relaxation kinetics of the original MSM for protein folding in a toy model mimicking an ideal  funnelled energy landscape. We compare the relaxation rates of the original model $\lambda_R$ with  1256 states with that of the renormalized effective model $\lambda'_R$ which only contains 2 states and we report on the calculated folding and unfolding rates $k_f$ and $k_u$. The symbols $\delta \lambda'_R$ and $\delta k$ represent, respectively, the relative error in the effective relaxation rate and our estimate of the theoretical uncertainty on the renormalized relaxation kinetic matrix elements.}
\label{tab:fun_kin_rates}
\end{table}

\section{Application to realistic systems}
\label{MDtest}
Finally, we discuss two applications of the RGC algorithm to characterize the structural dynamics of realistic polypeptide chains, based on the continuous- and discrete- time formalism, respectively.

\subsection{Continuous-Time MSM for the Conformational Dynamics of Alanine Dipeptide}
\begin{figure}[!h]
\includegraphics[width=14cm]{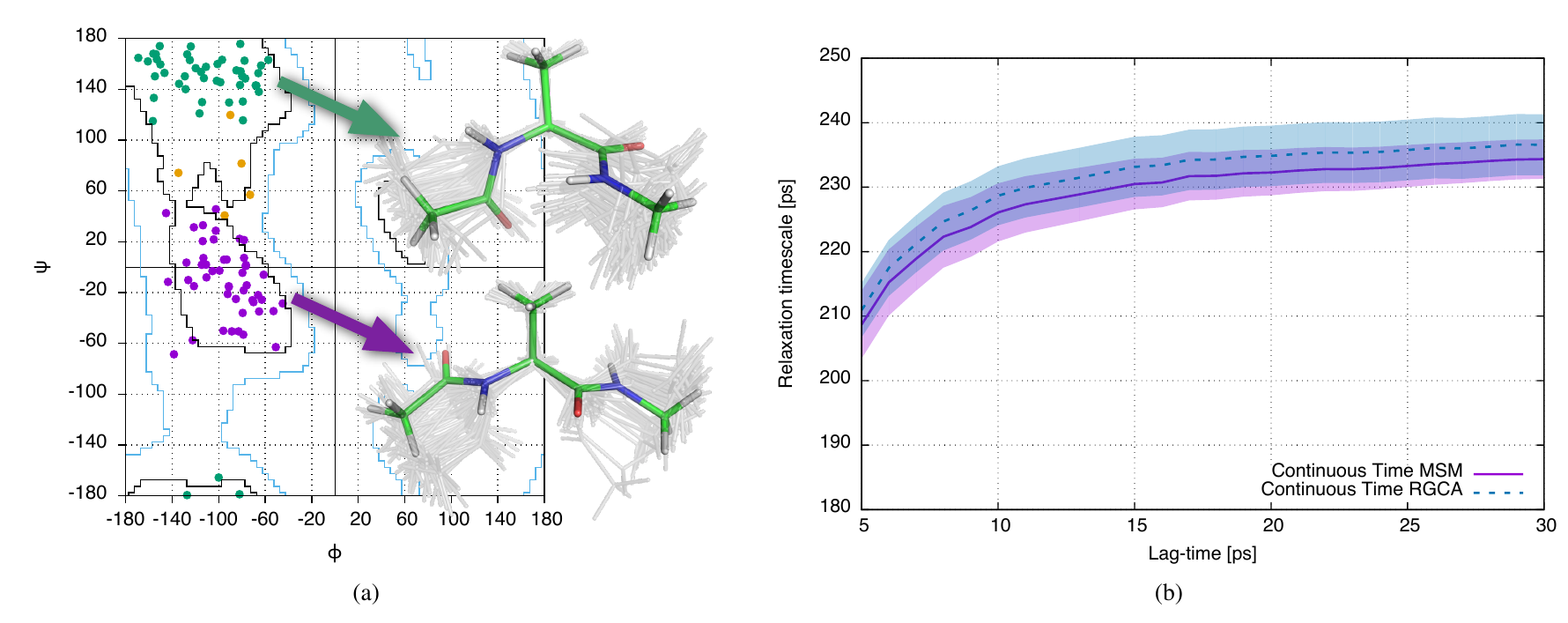}
\caption{ (a) Microstates of the continuous-time MSM for alanine dipeptide obtained using with $\tau=30ps$ plotted on the Ramachandran plot. The two macrostates identified by the RGC algorithm are represented with the set of green and purple dots and the figures in the inset show the corresponding configurations. In particular, the colored structures denote the most representative configurations  in the  macrostate, i.e. those  with the highest coefficient in the microstates basis expansion; (b) relaxation timescale in different approaches. The solid line represents the results of the original MSM with a standard error reported by the shaded red area. The prediction of our RG-based coarse-grained MSM is represented by the dashed line and its shaded region corresponds to the values $t_R \pm t_r\Delta$.}
\label{macro_info}
\end{figure}

As a first example, let us  consider alanine dipeptide, for which a set of  ten  10-ns long MD trajectories could be downloaded at the web-site \cite{link}. These MD simulations were performed using OpenMM 6.0.1 by integrating the Langevin equation at 300K with the AMBER99SB-ILDN force field in implicit solvent, with a  friction coefficient of 91/ps. Frames were saved every ps. 

In order to build a continuous-time MSM from these trajectories we used the MSMBuilder package \cite{MSMB}.
The 10 trajectories were initially featurized with respect to the dihedral angles, providing a 4-dimensional feature space. To identify a set of slow collective variables from linear combinations of these features we performed TICA with lag-time $\tau_{TICA}=$1~ps. We observed a clear decoupling between the first and the second TICA timescales and we retained only the 2 slowest independent components. Then, 100 discrete states were generated using the KMeans algorithm and, finally, a continuous-time MSMs was obtained according to the procedure described in Ref. \cite{MSM3, Kalb}: Namely,  a maximum-likehood estimator was applied to extract a rate matrix from the results of discrete-time MSMs with lag-times $\tau$ in the interval 5~ps$\le \tau\le 30$~ps. 
For all values of $\tau$ in this range,  we detected a gap between the first and the second relaxation timescales, of order $\Delta \equiv \frac{t_1}{t_R}~\simeq O(10^{-2})$. Finally, we applied our RGC algorithm to obtain a coarse-grained  MSM. 

In Fig.\ref{macro_info} we report the  100 microstates on the Ramachandran plot. Two macrostates where detected by our reduction procedure (represented with the set of green and purple dots), which clearly coincide with the $\alpha_R$ and $\beta/C5$ configurations of the  dipeptide.  The  molecular conformations in the microstates belonging to these two states are shown in the figures in the insets. The orange points denote the so-called overlapping states, i.e. microstates which could be assigned  to both  macrostates. We emphasize that a small overlap between macrostates  is expected if the gap in the relaxation timescale spectrum is not very large. Moreover, we note that the overlap is found in correspondence to the $C7_{\text{eq}}$ region, which was poorly sampled by the MD trajectory and therefore the RGC algorithm was not able to resolve it as a separate macrostate. These results show that the RGC algorithm provides physically sensible macrostates. 

Let us now discuss the  relaxation kinetics of the effective MSM and compare it to that of the original one. Throughout  the range lag-times used to define the rate matrix (see discussion above) our effective MSM reproduces the relaxation time of the original MSM to a very high accuracy, with a relative error always below $1\%$, consistently with a spectral gap $\Delta \sim 10^{-2}$. Such gap relative accuracy is also reported in Fig. \ref{macro_info}(b) by means of the blue shaded area, which represents the region $t_r \pm t_r \Delta$.

One important observation concerns the dependence of the relaxation times calculated from the original and reduced MSM on the lag-time, which is clearly evident in Fig. \ref{macro_info}. 
Ideally, one expects that relaxation times calculated from a MSM should be independent on the specific choice of lag-time  $\tau$. However, this is strictly valid  for MSM with infinite cut-off, i.e. with  $\Lambda\to \infty$. In practice, in building MSM from MD trajectories, a set of fast relaxation frequencies are projected out -- thus $\Lambda < \infty $ --- and some residual weak dependence on $\tau$ survives. This dependence is even more pronounced for renormalized MSM, in which additional modes are projected out. 
The logarithmic scaling with the cut-off is a well-understood feature of RG. Physically, it reflects the fact that MSMs are not as microscopic as the original MD description.

\subsection{Discrete-Time MSM for Native Dynamics of  a Realistic Protein} 

As a final example, we study the conformational dynamics in the native state of Bovine Pancreatic Trypsin Inhibitor (BPTI), starting from the $1$ms-long MD trajectory obtained by the DESRES group using the Anton supercomputer (all the technical information about this ultra-long trajectory can be found in the original publication \cite{Shaw}). 

To construct the MSM we adopted a procedure discussed in Ref. \cite{TICA3}, using the PyEmma package \cite{PyEMMA}: First, we down-sampled the trajectory in order to obtain 100000 frames separated by a timestep of $10$ns. Next, we  further reduced the size of the sample by representing the chain conformation using only the centers of the $C_\alpha$ atoms. After aligning the trajectory with respect to a target native structure, we used as features the $C_\alpha$ coordinates and used TICA to project onto the 2 slowest independent components using  $\tau_{TICA}=10~\mu$s. The slowest relaxation frequency,  which sets the UV cut-off of the MSM was found to be $\Lambda =0.15~\mu~\textrm{s}^{-1}$. 
We then used the KMeans algorithm to geometrically cluster the projected conformations into 100 clusters and obtained discrete-time MSM in a range of lag-times $6.8 ~\mu\text{s} \le \tau \le 7.5~\mu\text{s}$. We emphasize that the lag-time always satisfies the inequality $\tau > 1/\Lambda$, which is required for our MSM to be interpreted as a rigorous ET. 
A gap with $\Delta \simeq 0.25$ in the relaxation frequency spectrum of this MSM was detected, with 3 modes below the gap, implying the existence of three macrostates. 
Finally, we applied our RGC scheme to construct an effective MSM to describe the slow relaxation dynamics in the space of macrostates.

\begin{figure}[!h]
\includegraphics[width=14cm]{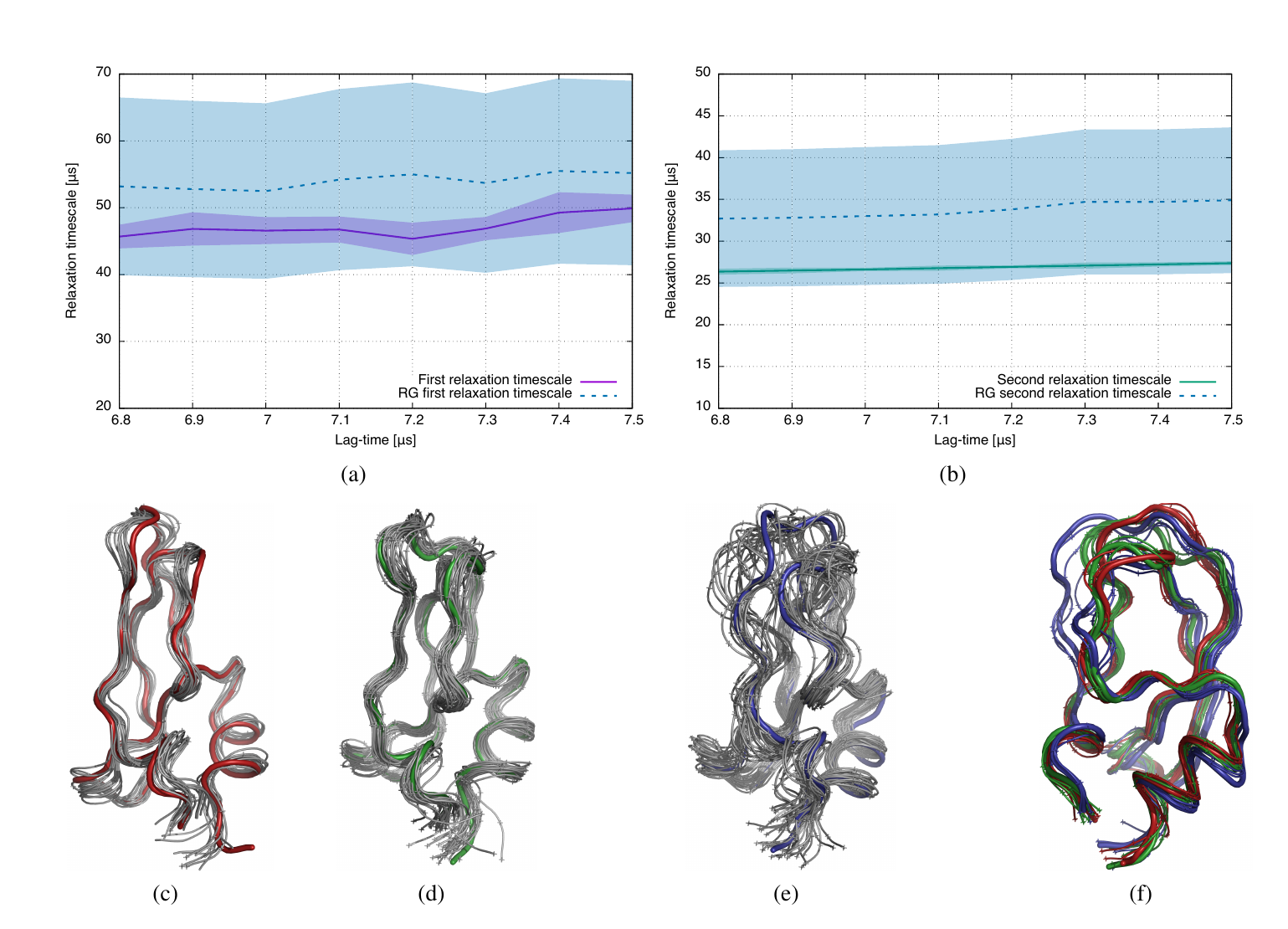}
\caption{(a)-(b) Lowest two relaxation timescales in different approaches. The solid lines represent the results of the original MSM with their standard error reported by the correspondent shaded areas. The predictions of our RG-based coarse-grained MSM are represented by the dashed lines and the shaded regions correspond to the values $t_i \pm t_i\Delta$ where $t_i$ is the i-th timescale; (c), (d), (e) show the three macrostates structures obtained using our RGC analysis, where the colored ribbons represent the most relevant structures in the three sets; (f) comparison of the five most relevant structures of the three macrostates.}
\label{graphs}
\end{figure}

In Fig.\ref{graphs}a we  compared results obtained in the original MSM with those calculated in the dimensionally reduced model.  Again, our method yields consistent results, predicting the original relaxation frequencies with a relative error of 20\%, which is in line with the expected accuracy, given the extent of the gap $\Delta$. As in the alanine dipeptide case, the shaded areas in Fig.\ref{graphs}a represent the  region $t_i \pm t_i \Delta$, estimating the uncertainty on the RGC predictions for the relaxation times.

In Fig.s \ref{graphs}c-\ref{graphs}e we show the structures of the three macrostates obtained using our RGC algorithm. The first two states are structurally very similar and show a compact conformation, while the third state is more flexible, especially in the upper loop region, which is outwardly folded. Fig.\ref{graphs}f shows the five more relevant conformations composing each of the three macrostates superposed together. These conformations are very similar to those predicted in a previous MSM analysis on BPTI dynamics, based on the of Robust Perron-Cluster Cluster Analysis (PCCA+) -- see \cite{TICA3} and references thereien--. This shows that the RGC scheme is able to correctly predict the metastable configurations, with accuracy comparable  with that of PCCA-based schemes.

\section{Conclusions}
\label{conclusions}
We have introduced  a rigorous approach based on the RG formalism to optimally cluster the microstates of a MSM  and compute the corresponding effective continuous- or discrete-time master equation.
Formulating MSMs as rigorous ETs  guarantees that the resulting effective MSM yields the same slow relaxation timescales of the original MSM, within an accuracy set by the spectral gap $\Delta$. When the spectrum contains multiple gaps, like in the first of the toy models we have presented,  the RGC can be used to  generate effective MSMs characterized by a different level of space and time resolution. 

All the tests we have made to assess the reliability of our algorithm showed that the method yields results which are in good agreement with the original model and theoretical expectations.  The computational cost of our dimensional reduction algorithm scales quadratically with the number microstates, and remains affordable for basically all  cases of practical interest. For example, the dimensional reduction of the MSM with 2500 microstates required  $\sim 20$s on a laptop computer.

We note that, at the algorithmic  level, our RGC scheme bears some similarities with the dimensional reduction scheme very recently proposed by Hummer and Szabo~\cite{HummerSzabo}. Indeed,  both methods are based on a projection formalism and give reduced models which accurately describe  the original kinetics. However, the  Hummer-Szabo method does not stand on the ET interpretation of MSM and does not require the Markovian approximation for the reduced model. Therefore it holds also in the absence of a gap in the relaxation frequency spectrum. On the other other hand,  it requires to postulate the number of macrostates and their structure  in terms of microstates,  while the present approach provides such a structure as a prediction, following directly from RG arguments.
The RGC algorithm displays also some analogy  with the approach developed by Gfeller and De Los Rios \cite{DeLosRios1, DeLosRios2}. The main difference is that, in the dimensional reduction scheme developed by these authors , the macrostates are constructed using a geometric  criterion (spatial proximity between microstates), rather than by  systematically exploiting the existence of a gap in the relaxation frequency spectrum.
Finally,  we note that our reduction scheme shares some common features also with the approach introduced by  Kube and Weber in Ref. ~\cite{pcca},  based on introducing the equivalent of  truncation and embedding operators. However,  in such a dimensional reduction method,  the macrostates are fuzzy and have to be identified  independently, using the PCCA+ analysis. Furthermore, Kube and Weber's method is not based on minimising the error introduced by the dimensional reduction.

\acknowledgments
We thank M. Harrigan, C. Hernandez and R. T. McGibbon for important help on the application of MSMBuilder, and G. Per\'ez-Hern\'andez for providing support with the PyEMMA package.   We acknowledge stimulating discussions with S. a Beccara and P. De Los Rios. Finally, we are grateful to  F. No\'e and A. Szabo for reading our manuscript and making important comments.

PF acknowledges special support from University of Trento through the grant ``Bando Progetti Strategici di Ateneo".

\newpage
\appendix
\begin{center}\Large{Appendix}
\end{center}

\section{States with Parametric Dependence Belong to $\mathcal{H}$}\label{app1}

In this appendix we show that if the distribution $\psi(x,t)$ (solution of the effective Schr\"odinger equation (\ref{SE}))  is expandable as a series of eigenstates of  the hermitian operator $H_{h}$ 
\begin{equation}
\psi(x,t) = \sum_{i=1}^N c_i~e^{-\lambda_i t}~ \phi_i(x)
\end{equation}
then, the corresponding state $|\psi(t)\ra$ can be expanded as linear combinations of microstates $\{|i\rangle\}$ (and $\{\langle i |\}$).

Indeed one finds for the ket-state:
\be
\begin{split}
|\psi(t)\rangle &=  \int dx ~ \sum_{j=1}^{N} ~c_j e^{-\lambda_j t}~ \phi_{j}(x)~|x\rangle \\
& = \sum_{j=1}^{N} c_j e^{-\lambda_j t} \int dx \sum_{i=1}^{N} ~C_{ji}^{-1}~ \pi^{(i)}(x)~ |x\rangle\\
&= \sum_{i=1}^{N} \sum_{j=1}^{N} ~c_j ~e^{-\lambda_j t} ~C_{ji}^{-1}~ |i\rangle 
\equiv \sum_{i=1}^{N} \frac{n_i(t)}{\sqrt{Z_i}}~|i\rangle
\end{split}
\ee
where $\pi^{(i)}(x)$ is the hermitian component of the state distribution $p^{(i)}(x)$.

\section{Pseudo-Inverse and its Representation in Terms of Singular Value Decomposion}\label{app3}

The pseudo-inversion generalizes the notion of matrix inversion to rectangular matrixes.
Namely, the pseudo-inverse $G^p$ of a matrix $G$ is defined by  the following properties (Moore-Penrose relationships).
\be
G G^P G = G, \quad G^P G G^P = G^P, \quad (G^P G)^\dagger = G^p G, \quad (G G^P)^\dagger = G G^P.
\ee

The pseudo-inverse of $G$ can be explicitly constructed using the singular-value decomposition (SVD):
\be
G=U~\Sigma~V^\dagger 
\ee
where $U$ is an $M \times M$ real or complex unitary matrix, $\Sigma$ is an $M \times N$ rectangular diagonal matrix with non-negative real numbers on the diagonal, and $V^\dagger$ is an $N \times N$ real or complex unitary matrix. The diagonal entries $\sigma_i$, of $\Sigma$ are  the singular values of $G$. 
 The $m$ columns of $U$ ($M$-dimensional vectors denoted with $ {\bf u}_1, \ldots  {\bf u}_M$) and the $N$ columns of $V$ ($N$-dimensional vectors denoted with $ {\bf v}_1, \ldots {\bf v}_N$) are the left-singular and right-singular vectors of $G$, respectively.


Namely,  in terms of singular value components, matrix $M\times N$ matrix  $G$ and its $N\times M$ pseudo-inverse matrix $G^P$ read 
\be
G_{I j} &=& \sum_{k=1}^M U_{I k} \sigma_k V^\dagger_{k j} \qquad (I= 1, \ldots, M, j=1, \ldots, N)\\
G^P_{i J} &=& \sum_{K=1}^M V_{i k} \frac{1}{\sigma_k} U^\dagger_{k J} \qquad (i= 1, \ldots, N, J=1, \ldots, M)\\
\ee
where $\sigma_K$ are the $M$ singular values. 

\section{Renormalisation of a discrete-time MSM}
\label{app6}
In this appendix, we extend the RG theory developed in section \ref{RSRG} to discrete-time MSMs. All definitions in Eqs. \eqref{truncation}, \eqref{psinv}, \eqref{proj_mat}, \eqref{projector} still hold in a stochastic system defined by a discrete master equation. 

Let us start by recalling that the microstates dynamics in the discrete-time case is expressed as
\begin{equation}
n_i(t+\tau) = \sum_{j=0}^N T_{ij}(\tau)~n_j(t) \;.
\end{equation}
Following Eq. \eqref{truncation}, the projection of the original $N$-dimensional population vector to an $M$-dimensional subspace leads to an effective lower-dimensional discrete-time master equation:
\begin{equation}\label{eff_dt}
n'_{I}(t+\tau) = \sum_{J=1}^MT'_{IJ}(\tau) n'_J(t) \qquad I,J = 1,\ldots,M<N.
\end{equation}
The matrix $T'(\tau)$ is the effective transition probability matrix and is given by
\begin{equation}
T'(\tau) = GT(\tau)G^P.
\end{equation}
Substituting the definitions of $n'_I$ and $T'(\tau)$ into \eqref{eff_dt} one finds
\begin{equation}
n_i(t+\tau) = \sum_{j=0}^N \left[ RT(\tau)R \right]_{ij}n_{j}(t)
\end{equation}
where we again defined the projector onto the relevant degrees of freedom $R = G^PG$. As in the continuous-time formulation we can compare the original representations of the dynamics with the one obtained embedding the effective description in the original $V^N$ space in order to minimize the difference between the two. To do this, we define the projector operator using Eq. \eqref{projector}  and approach the problem by supposing that we want to eliminate a single target vector from the Hilbert space:
\begin{equation}
\hat{R} = \hat{\Id}- |\textbf{v}\rangle \langle \textbf{v}| = \hat{\Id} - \hat{\mathcal{P}}_{|\textbf{v}\rangle}
\end{equation}
We have then:
\begin{equation}
n_i^{\text{original}}(t) - n_i^{\text{effective}}(t) = \sqrt{z_i} \langle i | \left( e^{-\hat{H}\tau} - \hat{R}e^{-\hat{H}\tau} \hat{R} \right) |\psi(t) \rangle = \langle i | \hat{\varepsilon} | \psi(t) \rangle
\end{equation}
where 
\begin{equation}
\hat{\varepsilon} =  \left\{ e^{-\hat{H}_{FP}\tau}, \hat{\mathcal{P}}_v \right\} - \hat{\mathcal{P}}_ve^{-\hat{H}_{FP}\tau}\hat{\mathcal{P}}_v
\end{equation}
represents the error operator. We have therefore to minimize the Frobenius norm of this operator: To this goal, let us compute its matrix elements in the microstate basis:
\begin{equation}
\varepsilon_{ij} = \sum_{k=1}^N\left( \Xi_{ik} v_kv_j + v_iv_k\Xi_{kj} - v_iv_j\sum_{l=1}^N v_k \Xi_{kl} v_l \right).
\end{equation}
Recalling the fact that $\Xi$ is symmetric, the Frobenius norm yields
\begin{equation}
\sum_{ij} |\varepsilon_{ij}|^2 = 2\sum_{i} \left(\sum_{k}\Xi_{ik}v_k\right)^2 - \Gamma^2 \qquad \Gamma \equiv \sum_{kl}v_k\Xi_{kl}v_l
\end{equation}
The minimization of the error norm with respect to the vector components $v_{\alpha}$ with the constraint that $\sum_{i}v_i^2=1$ finally provides the generalized eigenvalue equation
\begin{equation}\label{nl_eig_1v}
2 \sum_k \left( \Xi^2_{\alpha k} - \Gamma \Xi_{\alpha k} \right)v_k = \xi v_{\alpha} \qquad \Gamma = \sum_{ij} v_i \Xi_{ij} v_j
\end{equation}
where $\xi$ is a Lagrange multiplier enforcing the norm constraint. Even if Eq. \eqref{nl_eig_1v} is a non-linear eigenvalue problem, it admits a simple solution: let us rewrite Eq. \eqref{nl_eig_1v} using our operator formalism
\begin{equation}
2e^{-2\hat{H}\tau}|v\rangle - \langle v | e^{-\hat{H}\tau} | v \rangle e^{-\hat{H}\tau} |v\rangle = \xi |v\rangle
\end{equation}
and suppose that $\hat{H}|v\rangle = |\lambda| |v\rangle$ (which, in turn, means that $|v\rangle$ are also eigenstates of $e^{-\hat{H}\tau}$). We find that:
\begin{equation}
2 e^{-2\tau \hat{H}}|v\rangle - \langle v | e^{-\hat{H}\tau}|v\rangle e^{-\hat{H}\tau}|v\rangle = e^{-2|\lambda| \tau}|v\rangle
\end{equation}
Thus,  the vectors which minimize the error operator are  the eigenvectors of the operator $\hat{H}$, and  the error is given by
\begin{equation}\label{discrete_err}
|\hat{\varepsilon}|_F^2 = e^{-2|\lambda|\tau}
\end{equation}
We see that the error is globally minimized by projecting out all and only the eigenvectors with eigenvalues  above the gap. 

Let us now translate  the RGC algorithm  for the discrete time case. 
\begin{enumerate}
	\item compute the spectrum $\Lambda_T = \{\lambda_t\}_t$ of $T(\tau)$ up to $t=M$, where $\left|\log\lambda_M \right| \ll \left| \log\lambda_{M+1} \right|$, and the corresponding eigenvectors. Let us call $Z$ the eigenvector corresponding to $\lambda_1=1$ and define
	\begin{equation}\label{p_discrete}
	\Delta = \left| \frac{\log\lambda_M}{\log\lambda_{M+1}} \right|;
	\end{equation}
\item Compute the hamiltonian symmetrized transition probability matrix $\Xi_{ij}(\tau) = \sqrt{\frac{z_j}{z_i}}T_{ij}(\tau)$;
	\item Build the projector $R_{ij} = \sum_{i=1}^M v_i^{(K)}v_j^{(K)}$ where $v_i^{(K)}$ is the $i$-th components of the $K$-th lowest eigenstate of $\Xi(\tau)$;
	\item Extract the $M$ linearly independent vectors from $R$ to build the matrix $G$;
	\item The matrix $G$ may contain entries of order $\mathcal{O}\left(\frac{\lambda}{\Lambda}\right)$, some of which may be negative. Filter out all such  terms using the following rule. A matrix element $G_{Ij}$ is set to $0$ if there exists a $G_{Ik}$ element such that
	\begin{equation}
	\left| \frac{G_{Ij}}{G_{Ik}} \right| \leq \Delta
	\end{equation}
	Once the irrelevant terms have been deleted from the matrix, normalize its rows dividing them by $\sqrt{R_{ii}}$ (which is equivalent to normalizing the rows of the $G$ matrix to one);
	\item Solve the self-consistent equation $T'_{IJ}(\tau) = \sqrt{\frac{Z_I}{Z_J}} \sum_{i,j} G_{I i} \Xi_{ij}(\tau) G^T_{jJ}$ to obtain the reduced transition probability matrix;
	\item if $\lambda'_1 \not = 1$, redefine the $T'(\tau)$ spectrum $\Lambda'_T$ as
	\begin{equation}
	\Lambda''_T = \Lambda_T' - \Id\left( 1 - \lambda'_0 \right)
	\end{equation}
	and recompute the effective kinetic matrix as
	\begin{equation}
	T'(\tau) = V \Lambda_T'' V^{-1}
	\end{equation}
	where $V$ is the eigenstates matrix.
\end{enumerate}

\section{Properties of the Projection Matrix $R$}\label{app5}
Here we discuss some useful linear algebra results concerning the projection matrix $R$.
\begin{prop}\label{prop1}
Let $R$ be an $N\times N$ nihilpotent, singular and non-zero matrix with real coefficients. Then
\begin{equation}
R = R^P
\end{equation}
where $R^P$ is its Moore-Penrose pseudoinverse. 
\end{prop}\noindent
\textbf{Proof} Given the first Moore-Penrose relation applied to $R$ one has $R R^P R = R$. But since $R$ is nihilpotent one has $R = R^2 = R^3$, so that
\[
0 = R - R = RR^PR - RRR = R(R^PR - RR) = R(R^P-R)R
\]
which means that either $R=0$ (not possible by hypothesis) or $R^P=R$. $\Box$ \ \\

Before proving the next proposition, consider an $N \times N$ matrix $R$, with rank $M<N$. It is always possible to define a rectangular matrix, $\tilde{R}$, containing only the linearly independent rows of $R$, by defining an extraction matrix $E$ such that
\begin{equation}\label{RT}
\tilde{R} = ER
\end{equation}
In particular, this extraction matrix will be an $M \times N$ matrix containing zeros everywhere but the columns corresponding to the linearly independent rows, in which there is $1$. We will use such matrix in the following proposition.
\begin{prop}\label{lem:R_trasp}
Let $R$ be a rank deficient symmetric matrix showing also the properties listed in Lemma \ref{prop1} and let be $\tilde{R} = ER$ as defined in \eqref{RT}. Then
\begin{equation}
\tilde{R}^T = \tilde{R}^P
\end{equation}
\end{prop}\noindent
\textbf{Proof} The following chain of equivalences
\[
\tilde{R}^T = (ER)^T = R^TE^T = RE^T = R^PE^T = \tilde{R}^P
\]
closes the proof. $\Box$

\section{Implementation of the Fixed-Point Algorithm}\label{app7}

Our implementation of the fixed point method for the self-consistent calculation of the effective matrices is the following. Let us consider application to compute  the renormalized kinetic matrix (the case of the renormalized transition probability matrix is identical). By diagonalising an initial  guess matrix $K'_G$ one obtains the initial guess for the equilibrium eigenvector $Z^{(G)}$. One computes then an improved estimate for $K'_G$ using
\begin{equation}\label{self_cons}
\left(K'_{1}\right)_{KL} = (1-\alpha) \sqrt{\frac{Z_K^{(G)}}{Z^{(G)}_{L}}} \left(G H G^T \right)_{KL} + \alpha \left(K'_{G} \right)_{KL}
\end{equation}
where $ 0 \ll \alpha < 1$ is called the \emph{softening parameter}. This calculation has to be iterated until convergence is reached. Since the controlling parameter of the whole algorithm is the reduced equilibrium distribution $Z'$, then we can say that if
\begin{equation}\label{self_cons_conv}
\left|Z_i^{(N)} - Z_{i+1}^{(N)}\right| < \varepsilon \quad \forall i = 1,\ldots, M
\end{equation}
with fixed threshold $\varepsilon$, then convergence is reached and $K'$ is obtained as $K'_{N}$ where $N$ is the number of iterations. Another convergence criterion is to ask for the off-diagonal elements $k'_{ij}$ to converge. However this request is more expensive, since if $Z$ is described by $M$ elements, then the off-diagonal elements of $K'$ would be $M(M-1)$, requiring $M^2-2M$ more checks to assess convergence. The  two convergence requirements turn out to be  equivalent in all the practical cases we have considered.

In the calculations presented in this work we always choose as a guess the matrix
\begin{equation}
K'_G = G~H~G^T
\end{equation}
which has the advantage to yield the same spectrum of the converged matrix. Indeed, the normalization factor only acts on the non-diagonal elements of $K'$. 

\begin{figure}[!h]
\centering
\includegraphics[width=14cm]{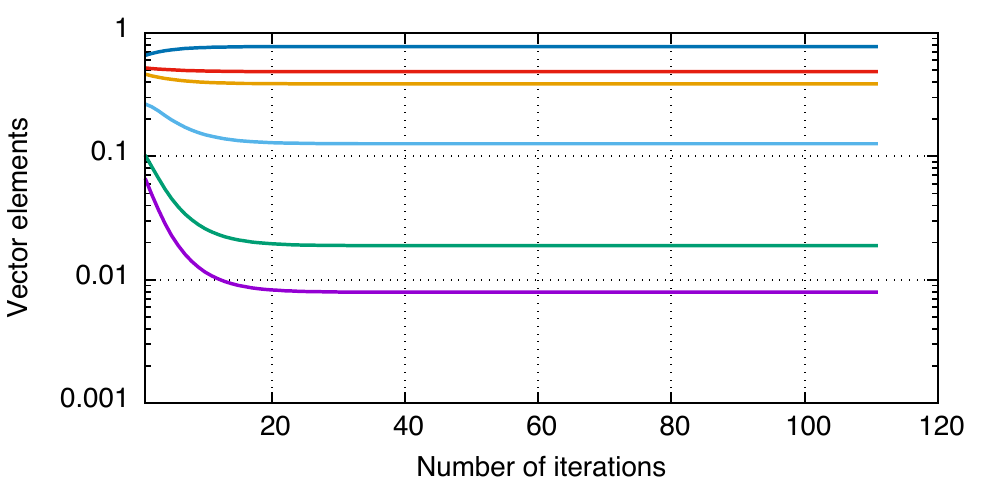}
\caption{Convergence of the six equilibrium vector components $Z'_i$, as a function of the number of self-consistent iterations.}
\label{convergence}
\end{figure}

As an example of application of the fixed-point algorithm, we show here its convergence properties when used to compute the effective kinetic matrix for the 6 macrostates system in section \ref{toy1}. In that case we choose $\varepsilon=10^{-12}$ and $\alpha=0.6$ and the algorithm converged to Eq. \eqref{app_K'1} in 112 steps - Fig. \ref{convergence}

\section{Step-by-step Solution of the Problem in section \ref{toy1}}\label{matrices}
This section simply provides a list of all the matrices employed in the step-by-step solution of the toy-model proposed in section \ref{toy1}.
\footnotesize
\be\label{app_K}
K=\left( 
\begin{array}{ccccccccccccc}
-2 & 1 & 0 & 0 & 0 & 0 & 0 & 0 & 0 & 0 & 0 & 0 & 0 \\
2 & -2 & 1 & 0 & 0 & 0 & 0 & 0 & 0 & 0 & 0 & 0 & 0 \\
0 & 1 & -1.02 & 0.01 & 0 & 0 & 0 & 0 & 0 & 0 & 0 & 0 & 0 \\
0 & 0 & 0.02 & -2.01 & 1 & 0 & 0 & 0 & 0 & 0 & 0 & 0 & 0 \\
0 & 0 & 0 & 2 & -1.02 & 0.01 & 0 & 0 & 0 & 0 & 0 & 0 & 0 \\
0 & 0 & 0 & 0 & 0.02 & -2.01 & 1 & 0 & 0 & 0 & 0 & 0 & 0 \\
0 & 0 & 0 & 0 & 0 & 2 & -2 & 1 & 0 & 0 & 0 & 0 & 0 \\
0 & 0 & 0 & 0 & 0 & 0 & 1 & -1.002 & 0.001 & 0 & 0 & 0 & 0 \\
0 & 0 & 0 & 0 & 0 & 0 & 0 & 0.002 & -2.001 & 1 & 0 & 0 & 0 \\
0 & 0 & 0 & 0 & 0 & 0 & 0 & 0 & 2 & -2 & 1 & 0 & 0 \\
0 & 0 & 0 & 0 & 0 & 0 & 0 & 0 & 0 & 1 & -1.02 & 0.01 & 0 \\
0 & 0 & 0 & 0 & 0 & 0 & 0 & 0 & 0 & 0 & 0.02 & -0.03 & 0.01 \\
0 & 0 & 0 & 0 & 0 & 0 & 0 & 0 & 0 & 0 & 0 & 0.02 & -0.01 \\
\end{array}
 \right)
\ee
\be\label{app_S}
H= -\left( 
\begin{array}{ccccccccccccc}
-2 & 1.41421 & 0 & 0 & 0 & 0 & 0 & 0 & 0 & 0 & 0 & 0 & 0 \\ 
1.41421 & -2 & 1 & 0 & 0 & 0 & 0 & 0 & 0 & 0 & 0 & 0 & 0 \\
0 & 1 & -1.02 & 0.01414 & 0 & 0 & 0 & 0 & 0 & 0 & 0 & 0 & 0 \\
0 & 0 & 0.01414 & -2.01 & 1.41421 & 0 & 0 & 0 & 0 & 0 & 0 & 0 & 0 \\
0 & 0 & 0 & 1.41421 & -1.02 & 0.01414 & 0 & 0 & 0 & 0 & 0 & 0 & 0 \\
0 & 0 & 0 & 0 & 0.01414 & -2.01 & 1.41421 & 0 & 0 & 0 & 0 & 0 & 0 \\
0 & 0 & 0 & 0 & 0 & 1.41421 & -2 & 1 & 0 & 0 & 0 & 0 & 0 \\
0 & 0 & 0 & 0 & 0 & 0 & 1 & -1.002 & 0.00141 & 0 & 0 & 0 & \\
0 & 0 & 0 & 0 & 0 & 0 & 0 & 0.00141 & -2.001 & 1.41421 & 0 & 0 & 0 \\
0 & 0 & 0 & 0 & 0 & 0 & 0 & 0 & 1.41421 & -2 & 1 & 0 & 0 \\
0 & 0 & 0 & 0 & 0 & 0 & 0 & 0 & 0 & 1 & -1.02 & 0.01414 & 0 \\
0 & 0 & 0 & 0 & 0 & 0 & 0 & 0 & 0 & 0 & 0.01414 & -0.03 & 0.01414 \\
0 & 0 & 0 & 0 & 0 & 0 & 0 & 0 & 0 & 0 & 0 & 0.01414 & -0.01 
\end{array}
\right)
\ee
\scriptsize
\be\label{app_R}
R =  \left( 
\begin{array}{ccccccccccccc}
0.2032 & 0.28623 & 0.28281 & -0.00046 & -0.00253 & -0.00001 & 0.00001 & 0.00003 & 0 & 0 & 0 & 0 & 0 \\
0.28623 & 0.40318 & 0.39836 & 0.00030 & -0.00221 & -0.00002 & 0 & 0.00002 & 0 & 0 & 0 & 0 & 0 \\
0.28281 & 0.39836 & 0.39363 & 0.00315 & 0.00183 & -0.00001 & -0.00001 & -0.00001 & 0 & 0 & 0 & 0 & 0 \\
-0.00046 & 0.00030 & 0.00315 & 0.33480 & 0.47190 & 0.00225 & -0.00062 & -0.00253 & 0 & 0 & 0 & 0 & 0 \\
-0.00253 & -0.00221 & 0.00183 & 0.47190 & 0.66515 & 0.00410 & 0.00046 & -0.00224 & 0 & 0 & 0 & 0 & 0 \\
-0.00001 & -0.00002 & -0.00001 & 0.00225 & 0.00410 & 0.19874 & 0.28204 & 0.28226 & 0.00008 & -0.00012 & -0.00023 &    0 & 0 \\
0.00001 & 0 & -0.00001 & -0.00062 & 0.00046 & 0.28204 & 0.40031 & 0.40064 & 0.00017 & -0.00008 & -0.00024 & 0 & 0 \\
0.00003 & 0.00002 & -0.00001 & -0.00253 & -0.00224 & 0.28226 & 0.40064 & 0.40098 & 0.00034 & 0.00016 & 0 & 0 & 0 \\    
0 & 0 & 0 & 0 & 0 & 0.00008 & 0.00017 & 0.00034 & 0.20308 & 0.28615 & 0.28274 & -0.00407 &  0.00005 \\
0 & 0 & 0 & 0 & 0 & -0.00012 & -0.00008 & 0.00016 & 0.28615 & 0.40321 & 0.39842 & -0.00284 & 0.00002 \\
0 & 0 & 0 & 0 & 0 & -0.00023 & -0.00024 & 0 & 0.28274 & 0.39842 & 0.39377 & 0.00579 & -0.00006 \\
0 & 0 & 0 & 0 & 0 & 0 & 0 & 0 & -0.00407 & -0.00284 & 0.00579 & 0.99994 & 0.00000 \\
0 & 0 & 0 & 0 & 0 & 0 & 0 & 0 & 0.00005 & 0.00002 & -0.00006 & 0 & 1.00000
\end{array}
\right)
\ee
\footnotesize
\be\label{app_R'}
R'= \left( 
\begin{array}{ccccccccccccc}
0.2032 & 0.28623 & 0.28281 & 0 & 0 & 0 & 0 & 0 & 0 & 0 & 0 & 0 & 0 \\
0.28623 & 0.40318 & 0.39836 & 0 & 0 & 0 & 0 & 0 & 0 & 0 & 0 & 0 & 0 \\
0.28281 & 0.39836 & 0.39363 & 0 & 0 & 0 & 0 & 0 & 0 & 0 & 0 & 0 & 0 \\
0 & 0 & 0 & 0.33480 & 0.47190 & 0 & 0 & 0 & 0 & 0 & 0 & 0 & 0 \\
0 & 0 & 0 & 0.47190 & 0.66515 & 0 & 0 & 0 & 0 & 0 & 0 & 0 & 0 \\
0 & 0 & 0 & 0 & 0 & 0.19874 & 0.28204 & 0.28226 & 0 & 0 & 0 & 0 & 0 \\
0 & 0 & 0 & 0 & 0 & 0.28204 & 0.40031 & 0.40064 & 0 & 0 & 0 & 0 & 0 \\
0 & 0 & 0 & 0 & 0 & 0.28226 & 0.40064 & 0.40098 & 0 & 0 & 0 & 0 & 0 \\    
0 & 0 & 0 & 0 & 0 & 0 & 0 & 0 & 0.20308 & 0.28615 & 0.28274 & 0 & 0 \\
0 & 0 & 0 & 0 & 0 & 0 & 0 & 0 & 0.28615 & 0.40321 & 0.39842 & 0 & 0 \\
0 & 0 & 0 & 0 & 0 & 0 & 0 & 0 & 0.28274 & 0.39842 & 0.39377 & 0 & 0 \\
0 & 0 & 0 & 0 & 0 & 0 & 0 & 0 & 0 & 0 & 0 & 0.99994 & 0 \\
0 & 0 & 0 & 0 & 0 & 0 & 0 & 0 & 0 & 0 & 0 & 0 & 1.00000
\end{array}
\right)
\ee
\be\label{app_G}
G = \tilde R = \left( 
\begin{array}{ccccccccccccc}
0.45079 & 0.63497 & 0.62738 & 0 & 0 & 0 & 0 & 0 & 0 & 0 & 0 & 0 & 0 \\
0 & 0 & 0 & 0.57864 & 0.81558 & 0 & 0 & 0 & 0 & 0 & 0 & 0 & 0 \\
0 & 0 & 0 & 0 & 0 & 0.44582 & 0.63270 & 0.63319 & 0 & 0 & 0 & 0 & 0 \\
0 & 0 & 0 & 0 & 0 & 0 & 0 & 0 & 0.45066 & 0.63501 & 0.62743 & 0 & 0 \\
0 & 0 & 0 & 0 & 0 & 0 & 0 & 0 & 0 & 0 & 0 & 1 & 0 \\
0 & 0 & 0 & 0 & 0 & 0 & 0 & 0 & 0 & 0 & 0 & 0 & 1 \\
\end{array}
\right)
\ee
\be\label{app_K'1}
K' = \left(
\begin{array}{cccccc}
-0.00794 & 0.00333 & 0 & 0 & 0 & 0 \\
0.00792 & -0.01666 & 0.00199 & 0 & 0 & 0 \\
0 & 0.01331 & -0.00280 & 0.00021 & 0 & 0 \\
0 & 0 & 0.00079 & -0.00814 & 0.00995 & 0 \\
0 & 0 & 0 & 0.00792 & -0.03000 & 0.00998 \\
0 & 0 & 0 & 0 & 0.02004 & -0.01000 \\
\end{array}
\right)
\ee
\be\label{app_K'2}
K' = \left(\begin{array}{cccccc}
-0.00792 & 0.00333 & 0 & 0 & 0 & 0 \\
0.00792 & -0.01664 & 0.00199 & 0 & 0 & 0 \\
0 & 0.01331 & -0.00278 & 0.00021 & 0 & 0 \\
0 & 0 & 0.00079 & -0.00812 & 0.00995 & 0 \\
0 & 0 & 0 & 0.00792 & -0.02998 & 0.00998 \\
0 & 0 & 0 & 0 & 0.02004 & -0.00998 \\
\end{array}
\right)
\ee
\be\label{app_G2}
G = \tilde R = \left( 
\begin{array}{ccccccccccccc}
0.11701 & 0.16543 & 0.16527 & 0.21472 & 0.30342 & 0.39880 & 0.56359 & 0.56283 & 0 & 0 & 0 & 0 & 0 \\
0 & 0 & 0 & 0 & 0 & 0 & 0 & 0 & 0.23925 & 0.33848 & 0.33860 & 0.48605 & 0.69086 \\
\end{array}
\right)
\ee
\be\label{app_K'3}
 K"= \left(\begin{array}{cc}
-6.45\times 10^{-4} & 5.6 \times 10^{-5} \\
6.45\times 10^{-4} &   -5.6 \times 10^{-5}
\end{array}
\right)
\ee

\begin{thebibliography}{99}


\bibitem{Pande0} V. Pande, K. Beauchamp, G. R. Bowman, Methods \textbf{52}, 2010, 99-105.
\bibitem{Noe0} G. R. Bowman, V. S. Pande, F. Noe, ''An Introduction to Markov State Models and Their Application to Long Timescale Molecular Simulation'', Springer (2014).
\bibitem{Prinz0} J.-H. Prinz, B. Keller, F. No\'e, Phys. Chem. Chem. Phys. \textbf{13}, 2011, 16912-16927.
\bibitem{Noe1} J.-H. Prinz \emph{et al.}, J. Chem. Phys. {\bf 134}, 2011, 174105.
\bibitem{MSMfund1} A. Bovier, M. Eckhoff, V. Gayrard, and M. Klein, Commun. Math. Phys. {\bf 228}, 2002, 219.
\bibitem{MSMfund2}  C. Sch\"{u}tte, C., A. Fischer, W. Huisinga, P. Deuflhard, J. Comput. Phys. \textbf{151}, 1999, 146-168.
\bibitem{MSMfund3} G. Biroli and J. Kurchan, Phys. Rev. {\bf E 64}, 2001.
\bibitem{MSMfund4} S. Tanase-Nicola and J. Kurchan, J. Stat. Phys. {\bf 116}, 1201, (2004).
\bibitem{Pande1} R. T. McGibbon, C. R. Schwantes, V. S. Pande, J. Phys. Chem. \textbf{B 118}, 2014, 6475.

\bibitem{Huang0} Y. Yao, R. Z. Cui, G. R. Bowman, D. A. Silva, J. Sun, X. Huang, J. Chem. Phys. \textbf{138}, 2013, 174106. 
\bibitem{Schutte0} P. Deuflhard, W. Huisinga, A. Fischer, C. Sch\"{u}tte, Linear Algebra Appl. 315, 39 (2000). 
\bibitem{Bowman0} G. R. Bowman, J. Chem. Phys. {\bf 137}, 134111 (2012). 
\bibitem{Bowman1} G. R. Bowman, L. Meng, X. Huang, J. Chem. Phys, {\bf 139}, 2013, 121905.
\bibitem{hidden} F. No\'e, H. Wu, J.-H. Prinz, N. Plattner, J. Chem. Phys. \textbf{139}, 2013, 184114. 
\bibitem{pcca} S. Kube, M. Weber, J. Chem. Phys. \textbf{126}, 2007, 024103. 
\bibitem{HummerSzabo} G. Hummer, A. Szabo, ''Optimal Dimensionality Reduction of Multistate Kinetic and Markov-State Models'', J. Phys. Chem., in press. 
\bibitem{DeLosRios1} D. Gfeller and P. De Los Rios, Phys. Rev. Lett. {\bf 99}, 2007, 038701.
\bibitem{DeLosRios2}D. Gfeller and P. De Los Rios, Phys. Rev. Lett. {\bf 100}, 2008, 174104.
\bibitem{Lepage} P. Lepage, ''How to Renormalize the Schr\"odinger equation'', lectures given at the VIII Jorge Andre Swieca Summer School (Brazil, 1997). ArXiv: nucl-th/9706029. 
\bibitem{RG1} A. Degenhard,  J. Rodriguez-Laguna, J. Stat. Phys. {\bf 106},  (2002).
\bibitem{RG2} A. Degenhard,  J. Rodriguez-Laguna, ``Renormalization Group Methods for Coarse-Graining of Evolution equations",  chapter in ''Model Reduction and Coarse-Graining Approaches for Multiscale Phenomena'',   Springer Berlin Heidelberg  (2006).
\bibitem{Sarich} Sarich et al., Multiscale Model. Simul., \textbf{8}, 1154-1177. 
\bibitem{Zwanzig0} R. Zwanzig, Phys. Rev. \textbf{124}, 1961, 983.
\bibitem{Mori} H. Mori, Prog. Theo. Phys., \textbf{33}, Issue 3, 1964, 423-455. 
\bibitem{TICA1} C. R. Schwantes, V. S. Pande, J. Chem. Theory Comput. \textbf{9} (4), 2013, 2000-2009.
\bibitem{TICA2} G. P\'{e}rez-Hern\'{a}ndez, F. Paul, T. Giorgino, G. De Fabritiis, F. No\'e, J. Chem. Phys. \textbf{139}, 015102, (2013).
\bibitem{TICA3} F. No\'{e}, C. Clementi, J. Chem. Theory Comput., 2015, \textbf{11} (10), pp 5002-5011.
\bibitem{TICA4} C. R. Schwantes, V. S. Pande, J. Chem. Theory Comput., \textbf{9},  2000 (2014).
\bibitem{PCA} R. T. McGibbon, C. R. Schwantes, V. S. Pande, J. Phys. Chem. B, 2014, \textbf{118} (24), pp 6475-6481.
\bibitem{Sriraman} S. Sriraman, I. G. Kevrekidis, G. Hummer, J. Phys. Chem. B, 2005, \textbf{109} (14), 6479-6484.
\bibitem{EL} J. Onuchic, Ann.l Rev. of Phys. Chem. {\bf 48}, 545 (1997).
\bibitem{link} \url{http://dx.doi.org/10.6084/m9.figshare.1026131}
\bibitem{MSMB} K. A. Beauchamp, G. R. Bowman, T. J. Lane, L. Maibaum, I. S. Haque, V. S. Pande, J. Chem. Theory Comput., {\bf 7}, 3412 (2011).
\bibitem{MSM3} R. T. McGibbon, V. S. Pande, ''Efficient maximum likelihood parameterization of continuous-time Markov processes'', in press 
\bibitem{Kalb} J. D. Kalbfleisch, J. F. Lawless, J. Am. Stat. Assoc. 80, 863, (1985).
\bibitem{Shaw} D. E. Shaw, P. Maragakis, K. Lindorff-Larsen, S. Piana, R. Dror, M. Eastwood, J. Bank, J. Jumper, J. Salmon, Y. Shan, W. Wriggers, Science \textbf{330}, 2010, 341-346.
\bibitem{PyEMMA} M. K. Scherer, B. Trendelkamp-Schroer, F. Paul, G. P\'erez-Hern\'andez, M. Hoffmann, N. Plattner, C. Wehmeyer, J.-H. Prinz, F. No\'e, J. Chem. Theo. Comput. \textbf{11}, 5525-5542, 2015. 


\end{thebibliography}
\end{document}